\documentclass{aa}  

\usepackage{graphicx}
\usepackage{txfonts}
\usepackage{lipsum}
\usepackage{amsmath}
\usepackage{enumitem}
\usepackage{natbib}
\usepackage[dvipsnames]{xcolor}

\usepackage{natbib,twoopt}
\usepackage[colorlinks,breaklinks=true]{hyperref}
\hypersetup{
    citecolor={blue},
    linkcolor={blue},
    urlcolor={blue}}

\newcommand{\di}{\mathrm{d}}
\newcommand{\masyr}{\, {\rm mas \, yr^{-1}}}
\NewDocumentCommand{\mul}{o}{%
  \ensuremath{\IfNoValueTF{#1}
    {\mu_\ell}
    {\mu_{\ell,#1}}}%
}
\NewDocumentCommand{\mub}{o}{%
  \ensuremath{\IfNoValueTF{#1}
    {\mu_b}
    {\mu_{b,#1}}}%
}
\NewDocumentCommand{\Sigmaobs}{o}{%
  \ensuremath{\IfNoValueTF{#1}
    {\Sigma_{\rm obs}}
    {\Sigma_{{\rm obs},\textit{#1}}}}%
}
\newcommand{\agama}{\textsc{agama}}

\newcommand{\Sigmatot}{\Sigma_{\rm{tot}}}
\usepackage{subcaption}         
\usepackage{lscape}             
\usepackage{placeins}           

\begin{document}

\authorrunning{Vasini et al.}
   \title{A Novel Approach to 3D Dust Mapping of the Central Molecular Zone}

\newcommand{\CLAP}    {Como Lake centre for AstroPhysics (CLAP), DiSAT, Universit{\`a} dell’Insubria, via Valleggio 11, 22100 Como, Italy}
\newcommand{\ZAHITA}  {Universit\"{a}t Heidelberg, Zentrum f\"{u}r Astronomie, Institut f\"{u}r Theoretische Astrophysik, Albert-Ueberle-Straße 2, D-69120 Heidelberg, Germany}
\newcommand{\UCL}   {Department of Physics and Astronomy, University College London, London WC1E 6BT, UK}
\newcommand{\UConn}   {Department of Physics, University of Connecticut, 196A Auditorium Road, Unit 3046, Storrs, CT 06269, USA}
\newcommand{\ISSA}{Institute of Space Sciences {\&} Astronomy, University of Malta, Msida MSD 2080, Malta}
\newcommand{\UF}{Department of Astronomy, University of Florida, Gainesville, FL 32611 USA}
\newcommand{\IAA}{Instituto de Astrofísica de Andalucía (CSIC), Glorieta de la
Astronomía s/n, E-18008 Granada, Spain}
\newcommand{\UCam}{DAMTP, Centre for Mathematical Sciences, University of Cambridge,  Wilberforce Road, Cambridge CB3 0WA,  UK}

\author{Arianna Vasini    \inst{\ref{CLAP}}\corrauth{arianna.vasini@uninsubria.it}
    \and
    Mattia~C.~Sormani   \inst{\ref{CLAP}                } \and
    Jason~L.~Sanders     \inst{\ref{UCL}                } 
    \and 
    Adam Ginsburg       \inst{\ref{UF}                  }
    \and
    Cara Battersby      \inst{\ref{UConn}               } 
    \and
    Francisco Nogueras-Lara  \inst{\ref{IAA}}             \and
    Rainer Sch\"odel      \inst{\ref{IAA}}
    \and
    Evangelia Deliporanidou \inst{\ref{UCam}}
    \and
    Marco Donati        \inst{\ref{CLAP}                } \and 
    Zi-Xuan Feng      \inst{\ref{CLAP} , \ref{ZAHITA}  } \and
    Karl Fiteni         \inst{\ref{CLAP}, \ref{ISSA}    }  \and 
    Savannah Gramze     \inst{\ref{UF}                  } \and 
    Xingchen Li         \inst{\ref{CLAP}                } \and 
    Dani~R.~Lipman      \inst{\ref{UConn}               } \and
    Perry~H.~Hatchfield \inst{\ref{UConn}} 
}

\institute{
            \CLAP       \label{CLAP}        \and            
            \UCL        \label{UCL}         \and 
            \UF         \label{UF}          \and
            \UConn      \label{UConn} 
            \and
            \IAA        \label{IAA}
            \and
            \UCam       \label{UCam}
            \and
            \ZAHITA     \label{ZAHITA} 
            \and
            \ISSA       \label{ISSA}
}

   \date{}

\abstract
   {
    The 3D distribution of dust and gas in the Milky Way's Central Molecular Zone (CMZ) is key to understanding gas inflows towardsthe Galactic Centre (GC), the process of star formation in this extreme environment, and the propagation of energetic cosmic rays originating from Sgr A*. However, while recent efforts have combined datasets in a Bayesian framework to estimate the near/far positions of individual molecular clouds in the CMZ, conflicts between different methodologies still remain and we are still lacking a comprehensive, model-independent map of all of the gas and dust in the CMZ, which is critical to address key science questions. Here we develop a new methodology to infer the 3D dust distribution of the CMZ.
    The key idea of the method is to use stellar proper motions to get probabilistic information about the unknown stellar distances through a model of the distribution of star positions and velocities of the nuclear stellar disc (NSD), co-spatial with the CMZ.
    Taking stellar proper motions and extinctions as input, the latter adopted as a proxy of the dust column density, the method returns the 3D dust distribution.  
    It is non-parametric, makes no assumptions on the functional form of the dust distribution, and is fundamentally distinct and largely independent of all existing methods. We show that the method can robustly and effectively reconstruct the mock 3D CMZ structure by testing it on a range of mock dust distributions, both analytically generated and taken from hydrodynamical simulations. Finally, we discuss the prospects for applying the method to real data.}

\titlerunning{A Novel Approach to 3D Dust Mapping of the Central Molecular Zone}
  

   \maketitle
\nolinenumbers

\section{Introduction}
\label{sec:intro}
The central region of the Milky Way, roughly within Galactocentric radius $R\lesssim$ 200 pc, hosts a ring-like structure of dense gas and dust known as the Central Molecular Zone (CMZ, \citealt{MorrisSerabyn96,Mills17,Henshaw+23}).
The CMZ is the analogue of the star-forming nuclear rings that are often found at the centre of external barred galaxies \citep[e.g.\,][]{Gleis2026}.
Its gas density, pressure, and temperature exceed those of the solar neighbourhood by orders of magnitude, making it a unique laboratory to study star formation in extreme environments (see \citealt{Henshaw+23} and references therein).

The gas distribution in the CMZ is asymmetric, with approximately 3/4 of the dense gas concentrated at positive longitudes \citep{Bally+88,Henshaw+23}. On the sky, the CMZ extends over a Galactic longitude range of $|\ell|\leq1.5^\circ$ and a latitude $|b|\leq1.0^\circ$.

Developing a three-dimensional model of the CMZ is key to understanding the inflow of gas towards the Circumnuclear Disc in the central few pc and ultimately onto Sgr A$^*$ \citep[e.g.\,][]{Lipman+26}, the star formation processes \citep{Henshaw+17} and the propagation of the cosmic rays in the Galactic centre \citep[e.g.\,][]{Scherer+22,Scherer+23,Obolentseva+24,Ren+25}. Moreover, a 3D extinction map of the CMZ would allow us to correct for selection effects in stellar observations and to further investigate the structure and the kinematics of the nuclear star cluster (NSC) and the nuclear stellar disc (NSD).

To tackle the CMZ three-dimensional gas distribution problem several strategies have been adopted over the years \citep[see][for a review]{Henshaw+23}. These include comparing emission and absorption ratios \citep{Sofue95,Sawada+04,Yan+17,Battersby+25a,Battersby+25b,Walker+25,Lipman+25} also with a Bayesian approach \citep{Lipman+26}, deprojecting $(l,b,v)$ datacubes using geometrical and dynamical models \citep{Molinari+11,Kruijssen+15,Ridley+17,Tress+20,Armillotta2020}, using X-ray echoes from molecular clouds \citep{Ponti+10,Terrier+18,Chuard+18,Stel+25}, using red clump stars as standard candles \citep{Zoccali2021,Nogueras+21b}, using stellar kinematics \citep{Chatzopoulos+15,Martinez-Arranz2022,Nogueras+26} and more recently, adopting machine learning on galaxy simulations \citep{Dubois+26}.

Despite these efforts, the exact geometry and the distance of several prominent molecular clouds remain uncertain (see \citealt{Henshaw+23,Mills+26}), with the overall consensus extending only so far as to suggest that the CMZ gas is organised into a toroidal structure.

In this paper we develop a new methodology to reconstruct the 3D distribution of gas in the CMZ. Our method builds upon the pioneering works of \citet{Chatzopoulos+15,Martinez-Arranz2022} and \citet{Nogueras+26}, expanding their approach into a systematic Bayesian framework (see Sect.~\ref{sec:method} for more details). Our method takes as input the stellar kinematics and extinction, and returns as output the 3D distribution of dust. It therefore provides constraints that are completely independent from those based on emission and absorption of spectral lines and deprojection of $(\ell,b,v)$ datacubes.

The paper is structured as follows. In Sect. \ref{sec:method} we present our method. In Sect. \ref{sec:application_mock} we test the performance of the method using mock dust profiles and mock stellar observations. In Sect. \ref{sec:discussion} we discuss the spatial resolution that can be achieved by our method and its limits and we investigate further the role of the proper motions in our framework. Finally, in Sect. \ref{sec:conclusion} we summarise our conclusions.

\section{Methodology}
\label{sec:method}

Our goal is to determine the 3D dust distribution in the CMZ by exploiting the  proper motions and the stellar extinctions of stars. Before developing the mathematical details, we briefly summarise the key ideas.

\begin{figure}
    \centering
    \includegraphics[width=1.0\linewidth]{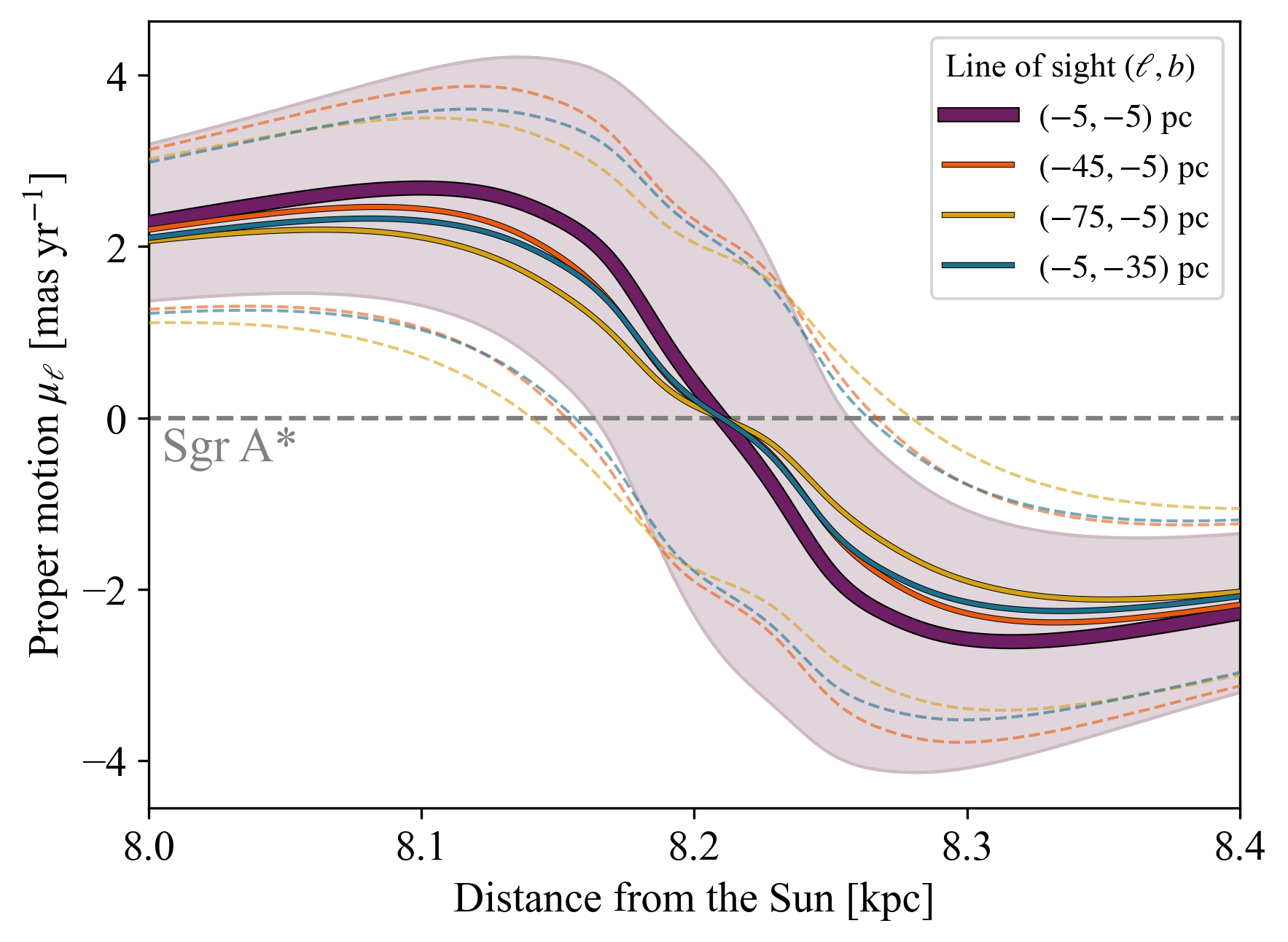}
    \caption{Proper motion parallel to the Galactic plane as a function of distance for the nuclear stellar disc stars in four different lines of sight computed adopting the dynamical model by \citet{Sormani+22}. The 1-$\sigma$ band for ($\ell,b$)=($-5$,$-5$) pc is denoted by the shaded area. For the other lines of sight the contours of the 1-$\sigma$ band are represented by the dashed lines.}
    \label{fig:proper_motion}
\end{figure}

Consider $N$ stars along the same line of sight $(\ell,b)$ with known extinctions and proper motions. For the moment we treat the line of sight as a pencil-beam (we will relax this assumption later). Our goal is to recover the dust density profile along this line of sight. Each star's extinction measures the dust column density in front of it, so the difference in extinction between two stars gives the amount of dust between them. If the stellar distances were known, the dust profile could be reconstructed immediately given enough stars. This is precisely the principle behind the 3D dust maps built with Gaia in the solar neighbourhood, where distances come from stellar parallaxes  \citep[e.g.][]{Lallement2019,Green2019,Edenhofer+24,Zucker2025}. Unfortunately, \textit{Gaia} does not reach the Galactic centre and getting parallaxes with sufficient precision to distinguish the near and far side of the CMZ is currently not possible. One of the key ideas of our methodology is that we can use the proper motions of the stars as a proxy of distance (see Fig. \ref{fig:proper_motion}).

The dust and gas in the CMZ reside within the nuclear stellar disc (NSD), a flattened stellar structure that dominates the Milky Way's stellar density at $R<300$ pc. Crucially, the NSD rotates, with the near (far) side moving towards positive (negative) longitude. For the line of sight through the GC, the difference between the average proper motion in the direction parallel to the Galactic plane of a star in the near side and a star in the far side is approximately $5 \masyr$ \citep{Schonrich+15,Nogueras22}, which is well within reach of current surveys, and the $1$-$\sigma$ dispersion for a star on either side is about $1 \masyr$ (see shaded band in Fig.~\ref{fig:proper_motion}). Thus, knowing the proper motion of a star we can place it confidently on the far or near side, which in turn constrains the position of the dust in front of it. This principle was exploited by \citet{Martinez-Arranz2022} and \citet{Nogueras+26} to determine the position of selected molecular clouds by looking at jumps in extinction as a function of proper motions. In this paper, we develop this idea into a fully Bayesian framework that exploits the information systematically.

\subsection{Mathematical framework}

\subsubsection{General formulas}
Since the CMZ area on the sky plane is relatively small we can adopt a Cartesian coordinate system centred on the GC, with the y-axis pointing along the line of sight, the z-axis perpendicular to the Galactic plane and the x-axis perpendicular to the other two to make it a right-handed Cartesian system.
Consider a line of sight beam of finite cross section centred on $(\ell,b)$ and a sample of $N$ stars along it, for which we have measured the dataset $D$.
Let us denote with $\Theta$ the parameters of our model, which in our case will be the dust profile that we want to recover. According to Bayes' theorem: 

\begin{equation}
    P(\Theta|D) = P(D|\Theta) \frac{P(\Theta)}{P(D)} \,,
    \label{eq:bayes}
\end{equation}

where $P(\Theta|D)$ is the probability of the parameters $\Theta$ given the observed data $D$ (which following standard nomenclature we will refer to as the posterior), $P(D|\Theta)$ is the probability of obtaining the data $D$ assuming a particular set of parameters $\Theta$ (which we will refer to as the likelihood), $P(\Theta)$ is the prior on the parameters $\Theta$, and $P(D)$ is the probability of the data, which acts as a normalisation constant ensuring that the posterior integrates to unity, $\int P(\Theta|D)\,\di\Theta = 1$. Under these assumptions Eq.~\eqref{eq:bayes} reduces to:
\begin{equation}
     P(\Theta|D) \propto P(D|\Theta)P(\Theta)   \,.  \label{eq:bayes2}
\end{equation}

Thus, Bayes theorem allows us to find the probability of a certain dust configuration given our data, $P(\Theta|D)$, by calculating the much easier probability of our data given the dust distribution, $P(D|\Theta)$. We now specify how to calculate the right-hand side of Eq.~\eqref{eq:bayes2}.

Let us define $\Sigma(y)=\int_{y_{\rm min}}^y \rho(y') \di y'$ as the average column density within our beam up to distance $y$, obtained by integrating the cross-section-averaged density $\rho(y)$. In the notation above $\Sigma(y)$ corresponds to $\Theta$. We do not introduce any a priori assumption on the shape of the profile $\rho(y)$, and therefore on $\Sigma(y)$, meaning that we assume the prior $P(\Theta)$ to be uniform. Eq.~\eqref{eq:bayes2} can be rewritten as:
\begin{equation}
    P(\Theta|D) \propto P(D|\Theta) \,.  \label{eq:bayes3}
\end{equation}
We now define the likelihood as:
\begin{equation}
    P(D | \Theta) = \mathcal{L}(D |\Sigma)
    \label{eq:P=L}
\end{equation}
and assuming that each star is independent, Eq.~\eqref{eq:P=L} can be rewritten as:
\begin{equation}
  \mathcal{L}(D |\Sigma)= \prod_{j=1}^N \mathcal{L}_j (D_j |\Sigma) \,.
  \label{eq:L_indepstars}
\end{equation}
The quantity $\mathcal{L}_j (D_j | \Sigma)$ is the probability that we measure the set of observables $D_j$ for the star $j$, given that the true column density profile along the line of sight is given by $\Sigma(y)$. Using the basic rules of probability we can write this as:
\begin{equation}
\boxed{
\mathcal{L}_j ( D_j |\Sigma) = \int P(D_j|y,\Sigma) P(y|\Sigma) \di y\,.}
    \label{eq:L=int}
\end{equation}

Eq.~\eqref{eq:L=int} is the most general version of our likelihood for a generic dataset $D$. We now specify Eq.~\eqref{eq:L=int} to the case where $D$ is composed of stellar extinctions $A_j$ and proper motions parallel to the Galactic plane $\mul$ (generalisations to cases in which we have other additional observations such as line-of-sight velocities $v_{\rm los}$, proper motions $\mub$ or other stellar parameters are straightforward):
\begin{equation}
    D=\{ (\mul[j],A_j) \} \qquad \text{with} \qquad j=1,\ldots,N \,.
    \label{eq:dataset}
\end{equation}
Given Eq.~\eqref{eq:dataset} we can rewrite Eq.~\eqref{eq:L=int} as:
\begin{align}
 \mathcal{L}_j (\mul[j],A_j |\Sigma) & = \int P(\mul[j],A_j|y,\Sigma) P(y|\Sigma) \di y \\
 & = \int P(\mul[j]|y) P(A_j|\Sigma ,y) P(y|\Sigma) \di y
 \label{eq:likelihood1}
\end{align}
where in going from the first to the second line we have assumed that $\mul$ depends only on the distance $y$ (which is reasonable because the column density will not affect two stars with different proper motions differently) and that $A_j$ depends only on the true column density $\Sigma(y)$ at distance $y$ (which is reasonable because the proper motion or the distance will not affect the observed extinction of two stars with the same true column density $\Sigma$). 

Equations \eqref{eq:L_indepstars} and \eqref{eq:likelihood1} are the formulas that constitute the basis of our method. Now we specify how we calculate the single terms in Eq.~\eqref{eq:likelihood1}.

\subsubsection{$P(\mul[j]|y)$}
\label{sec:P_mu}

To calculate the probability $P(\mul[j]|y)$ that the $j$-th star has a certain proper motion $\mul[j]$ given its distance, we need a model of the stellar kinematics as a function of distance. Since the goal of this paper is to provide a proof of concept of the method, we assume that the only stellar component is the NSD, and we adopt the axisymmetric dynamical model of \citet{Sormani+22} to compute the proper motion distribution of stars as a function of distance. More realistic configurations that include additional components, such as the NSC \citep{Vasiliev+26}, the large-scale bar \citep{Portail2017}, the Galactic disc \citep{Binney2023}, and possibly a nuclear bar \citep{Fiteni+26b}, can be added but for simplicity we leave this to a future paper.

Hereafter we also substitute $\mul[j]$ with the stellar velocity $v_{\ell,j}$. Since the NSD is small compared to its distance from the Sun (radius $R\simeq 100$~pc, distance $\sim 8.2$~kpc as in \citealt{Sormani+22}), the proper motion  $\mul[j] = v_{\ell,j} / d_j$ of a star in the front and a star in the back of the NSD changes by only $\sim 2\%$ due to the distance, less than the typical observational error on $\mul$, and we can consider $v_\ell$ and $\mul$ to be essentially equivalent.

The likelihood can then be rewritten as:
\begin{equation}
    \mathcal{L}_j (v_{\ell,j},A_j | \Sigma) = \int P(v_{\ell,j}|y) P(A_j|\Sigma,y) P(y|\Sigma) \di y \,.
    \label{eq:likelihood_v}
\end{equation}

For simplicity, we have so far neglected the uncertainty on $v_{\ell}$, that can however be incorporated in our framework as we show in Appendix~\ref{sec:error_vl}. 

\subsubsection{$P(y|\Sigma)$}
\label{sec:P_y}

This gives the probability of observing a star at distance $y$ given the column density profile $\Sigma(y)$. In general, it will depend both on the stellar density as a function of distance $P(y)$  and on the selection function $ S(y|\Sigma)$ of the survey:
\begin{equation}
    P(y|\Sigma) = P(y)S(y|\Sigma),
    \label{eq:P_y_Sigma}
\end{equation}
where $S(y|\Sigma)$ is the probability that a star at distance $y$ ends up in our survey given the column density $\Sigma$. 

In general, $S(y|\Sigma)$ can be obtained as:
\begin{equation}
    S(y|\Sigma)=S_{\rm{post}}(y|\Sigma)\cdot\int^{+\infty}_{-\infty} C(m(M,y,\Sigma)) \phi(M|y)\di M.
    \label{eq:sel_function_general}
\end{equation}
$\phi(M|y)$ is the luminosity function of the stellar population defined as the fraction of stars at distance $y$ with absolute magnitude between $M$ and $M+\di M$. It is normalised so that $\int\phi(M|y)\di M=1$ for every $y$ and accounts for age and metallicity gradients along the line of sight. $C(m(M,y,\Sigma))$ is the completeness function of the survey where $m(M,y,\Sigma)=M+\rm{DM}(y)+A(y|\Sigma)$ is the apparent magnitude of a star of absolute magnitude $M$ at distance $y$, corrected for the distance modulus $\rm{DM}(y)$ and the line-of-sight extinction $A(y|\Sigma)$ and $S_{\rm{post}}(y|\Sigma)$ accounts for any further selection applied. $C(m(M,y,\Sigma))$ captures all the selection effects introduced by the survey, including the magnitude limit, the seeing, the crowding and the saturation of the brightest stars. $S_{\rm{post}}(y|\Sigma)$ on the other hand, includes further selections such as the allocation of the spectroscopic fibres, the availability of the proper motions measurements and colour cuts.  
In general, given the characteristics of the survey one can construct the appropriate $S(y|\Sigma)$.

To perform the tests presented in the paper for simplicity we set the selection function to unity, so that 
\begin{equation}
\boxed{
P(y|\Sigma) = P(y) \,.}
\label{eq:P_y}
\end{equation}
We relax this assumption in Sect.~\ref{sec:no_proper_motions} to investigate the role of the proper motions in our framework.
We calculate $P(y)$ as the 3D stellar density in the NSD model of \citet{Sormani+22}.

\subsubsection{$P(A_j|\Sigma,y)$}

This term is the probability of observing a certain extinction $A_j$ given that the real column density is $\Sigma$. To keep the derivation and the mock data tests as simple as possible, we adopt two simplifying assumptions; we assume that the extinction $A_j$ and the column density towards the star, denoted by $\Sigmaobs[j]$, are linearly related by $ A_j = k\, \Sigmaobs[j]$ with constant $k$. 
Given this linear relation, in our tests we can use directly $\Sigmaobs[j]$ instead of $A_j$ without introducing any bias. Under this assumption and the assumption in the previous section that $P(y|\Sigma)=P(y)$, Eq.~\eqref{eq:likelihood_v} can be rewritten as:
\begin{equation}
\boxed{
\mathcal{L}_j (v_{\ell,j},\Sigmaobs[j] | \Sigma)  = \int P(v_{\ell,j}|y) P(\Sigmaobs[j]|\Sigma,y) P(y) \di y\,.}
 \label{eq:likelihood_sigma}
\end{equation}
We model $P(\Sigmaobs[j]|\Sigma,y)$ as a Gaussian:
\begin{equation}
\boxed{
P(\Sigmaobs[j]|\Sigma,y) = \frac{1}{\sqrt{2\pi}\sigma_j}\text{exp}\Bigg(-\frac{(\Sigma(y)-\Sigmaobs[j])^2}{2\sigma_j^2}\Bigg)\,,}
    \label{eq:p_sigma}
\end{equation}
where
\begin{equation}
    \sigma_j = \sqrt{a^2\sigma_{\rm{fov}}^2+\sigma_{\rm{obs,j}}^2} \,,
    \label{eq:sigmaj_real}
\end{equation}
incorporates two effects. The $\sigma_{\rm{obs,j}}$ is the observational error on $\Sigmaobs[j]$ that we set equal to $0.02\,\Sigmatot$ after calibration. The $\sigma_{\rm{fov}}$ is the standard deviation of extinction across the beam section. It takes into account that the observed extinction can fluctuate across a finite cross-section, and would be zero for a pencil-beam line of sight. $\sigma_{\rm{fov}}$ is distribution-dependent: in the tests presented below, it ranges from a few per cent of $\Sigmatot$ for the smoothest profiles to approximately $0.5\,\Sigmatot$ for the most irregular ones. The $a$ is a dimensionless scaling factor that we set to $0.1$ after mock data calibration.

\subsubsection{Set up and numerical implementation}
\label{sec:algorithm_implementation}

We divide the sky plane in independent fields of view each covering an area of 10 pc $\times$ 10 pc, similar to the size of the KMOS fields of view in the survey of \citet{Fritz+21}. We assume the Galactic centre to be at a distance $y_{\rm{GC}}=8.2$~kpc from us \citep{GravityCollab+21}.
We then parameterise the column gas density $\Sigma$ by dividing the line of sight into $M$ identical bins and assigning a column density $\Sigma_i$ to each bin. The $\Sigma$ is then represented by the set:

\begin{equation}
    \Sigma = \{\Sigma_1,...,\Sigma_M\}.
    \label{eq:sigma_parameters}
\end{equation}
To perform the marginalisation over $y$ in Eq.~\eqref{eq:likelihood_sigma} we discretised it on a regular grid of positions $y$ assumed for simplicity to be the $M+1$ bin edges. We exclude any foreground contribution.
In this convention, the column gas density in front of a star at distance $y$ is computed as:
\begin{equation}
    \Sigma(y)=\sum_{i=1}^M \Sigma_i \cdot w_i(y)
\end{equation}
where $w_i(y)=1$ if the $i^{th}$ bin is in front of $y$ and $w_i(y)=0$ if it is behind. This approach guarantees that no assumption on $\Sigma(y)$ is introduced and that the prior $P(\Theta)$ is actually uniform as assumed in Eq.~\eqref{eq:bayes3}. 
We also assume that the total column density on the line of sight, $\Sigmatot$, is known from observations (in real cases where the line of sight is very extincted, one might obtain this information from  molecular spectral line measurements).

For every line of sight we run a Monte Carlo Markov Chain (MCMC) using the likelihood in Eq. \eqref{eq:likelihood_sigma}. We use the No-U-Turn Sampler (NUTS) algorithm implemented in the \textsc{numpyro} package \citep{Numpyro19,Numpyro2_19} with 4 chains and 1500 steps. We do not run the MCMC directly on the $\Sigma_i$ parameters, but we remap the $\Sigma_i$ as $\vartheta_i$ by using a \emph{softmax} transformation:
\begin{equation}
\Sigma_i({\vartheta_1,...\vartheta_M}) = \Sigma_{\rm{tot}} \cdot \frac{e^{\vartheta_i}}{\sum_{k=1}^{M} e^{\vartheta_k}} \,.
\label{eq:softmax}
\end{equation}
This has the advantage that $\sum_k^M \Sigma_k=\Sigmatot$ 
and $\Sigma_k > 0\,\,\forall k$ are automatically verified, so we do not need to impose them as separate constraints. Under this transformation the likelihood becomes:
\begin{equation}
\mathcal{L}(v_{\ell},\Sigmaobs|\vartheta) = \mathcal{L}(v_{\ell},\Sigmaobs|\Sigma(\vartheta)) \cdot |\rm{det}\,J|
\label{eq:like_vartheta}
\end{equation}
where $|\rm{det}\,J|$ is the determinant of the Jacobian of the softmax transformation:
\begin{equation}
|\text{det}\,J|=\frac{\Sigmatot^{M-1}\prod_i^Me^{\vartheta_i}}{\Big(\sum_i^M e^{\vartheta_i}\Big)^M}\,. 
    \label{eq:jacobian}
\end{equation}

Once the MCMC has explored the parameter space, it returns S posterior samples $\vartheta^{(1)},...,\vartheta^{(S)}$ each of which, through Eq.~\eqref{eq:softmax}, corresponds to a dust profile  $\Sigma^{(s)}=(\Sigma^{(s)}_1,...,\Sigma^{(s)}_M)\,\in\mathbb{R}_+^M$. 

The reconstructed dust distribution is then given by the geometric median
restricted to the $S$ MCMC samples, $\Sigma^{(1)}, \dots, \Sigma^{(S)}$,
defined as the sample given by
$\hat{\Sigma} = \arg\min_{\Sigma \in \{\Sigma^{(1)}, \dots, \Sigma^{(S)}\}} \sum_{s=1}^{S} \lVert \Sigma - \Sigma^{(s)} \rVert_2$.

The code is publicly available at \url{https://github.com/AriannaVasini/3D_CMZ_dust_tomography}.

\begin{figure*}
    \centering
    \includegraphics[width=1\linewidth]{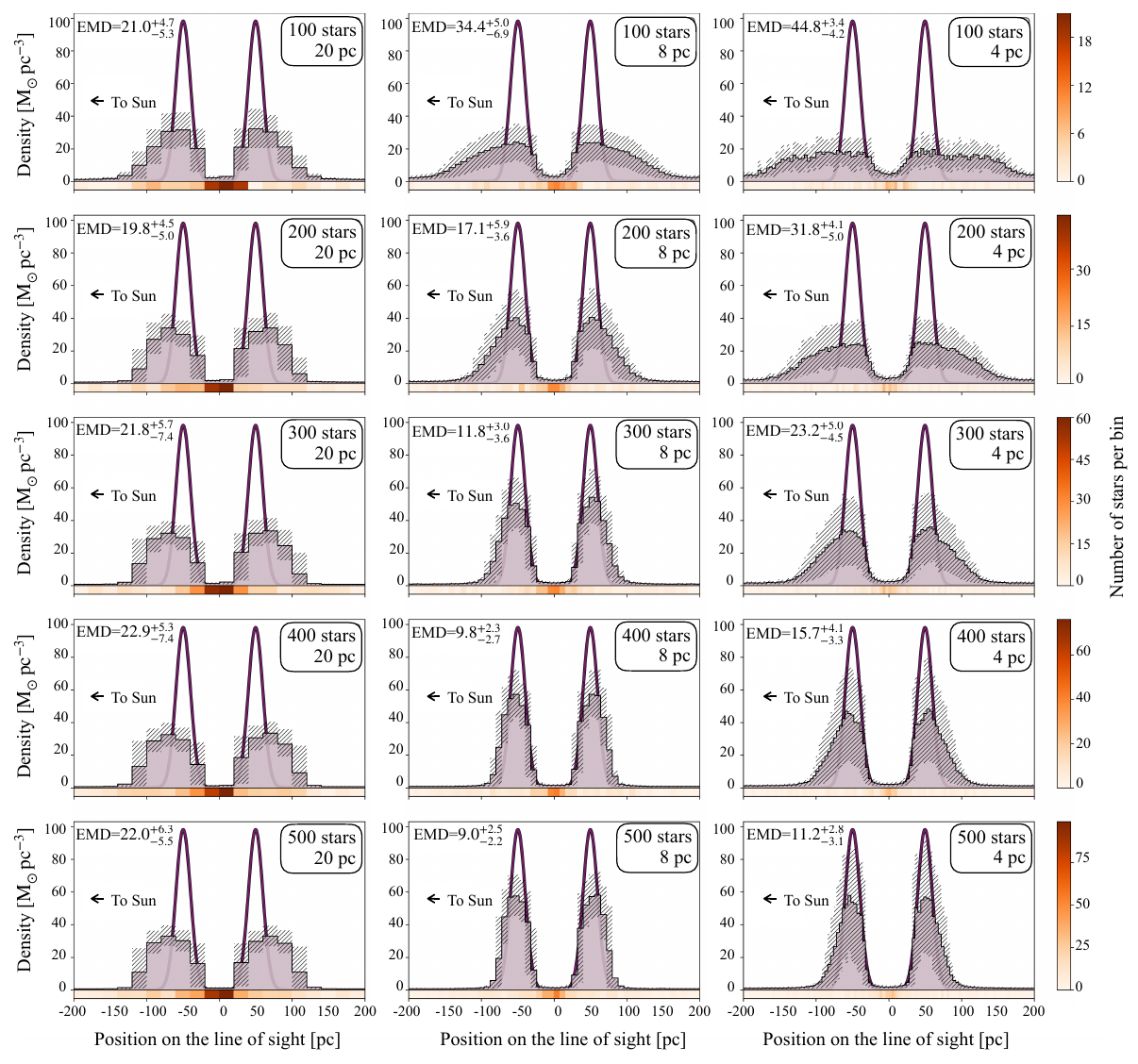}
    \caption{Recovered gas profile (lilac histograms) averaged over 500 realisations with posterior uncertainties (shaded area) compared to the mock gas profile Ring50 (purple solid line) along ($\ell,b$)=($-$5,$-$5) pc for different numbers of stars and different bin sizes, as indicated in the upper right of each panel. In the upper left we report the median EMD expressed in pc. The x axis represents the line of sight where 0 is the orthogonal of the Galactic centre on the line of sight and the Sun is located at $y=-8.2$ kpc. The orange bar below the panels shows the histogram of the star positions on the line of sight binned according to the adopted resolution.}
    \label{fig:megaplot}
\end{figure*}

\begin{figure}
    \centering
    \includegraphics[width=1.0\linewidth]{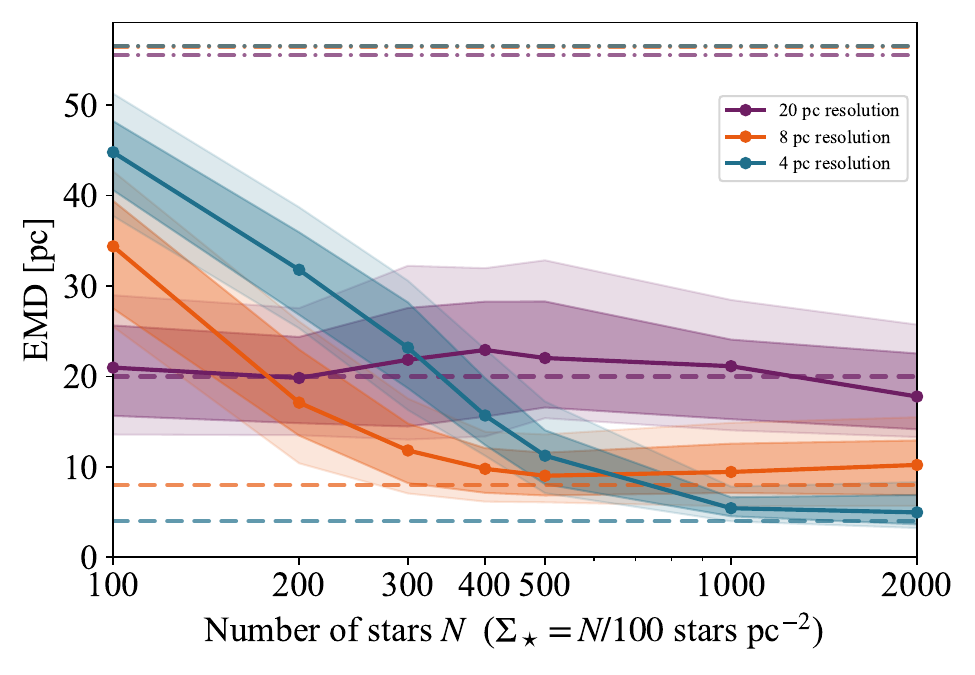}
    \caption{Median EMD over the 500 realisations as a function of the number of stars $N$ in the catalogue for the three resolutions tested. Given the $10 \times 10$ pc size of the fields, the $x$-axis can equivalently be read as stellar number surface density, $\Sigma_\star = N/100$ stars pc$^{-2}$. The shaded areas represent the 68\% and 90\% confidence intervals. The dashed lines mark the resolutions adopted and the dashed-dotted lines mark the EMD for a flat profile.}
    \label{fig:emd_vs_nstars}
\end{figure}

\section{Application to mock datasets}
\label{sec:application_mock}
We tested the performance of the method on different mock dust distributions and mock stellar catalogues. We generated a sample of mock NSD stars with 3D position and velocities by means of the \agama\ library \citep{Vasiliev19} assuming no uncertainties, and various mock gas distributions as described in the following section.

\subsection{Mock gas distributions}
\label{sec:mock_data}
We tested five mock gas distributions. For simplicity in every model we assume that all the dust responsible for stellar extinction lies within the CMZ, i.e. we neglect any foreground contribution from Galactic disc clouds. Model Ring50 is a smooth toroidal distribution described by the equation:
\begin{equation}
    \rho_{R}(R,z) = \rho_{0}\,\, \rm{exp}\,\Bigg(\frac{-(\textit{R} - \textit{R}_\textit{R})^2}{2\textit{W}_\textit{R}^2}\Bigg)\,\rm{exp}\,\Bigg(\frac{-\textit{z}^2}{2\textit{H}_\textit{R}^2}\Bigg)
    \label{eq:Ring}
\end{equation}
where $(x,y,z)$ are Cartesian coordinates centred on the Galactic centre, $\textit{R}=\sqrt{x^2+y^2}$, $\textit{R}_\textit{R}$, $\textit{W}_\textit{R}$ and $\textit{H}_\textit{R}$ are the scale radius, the scale width and the scale height of the ring and $\rho_0$ is tuned to reproduce the total mass $M_{\rm CMZ}=2\times 10^7\,M_{\odot}$ \citep{Dahmen+98}.
The values of the parameters are listed in Table \ref{tab:models}.

\begin{table}
\centering
\begin{tabular}{|c|c|c|c|c|}
  \hline
   & $R_R$  & $W_R$  & $H_R$  & $\alpha$ \\
  & [$\rm pc$]  & [$\rm pc$]  & [$\rm pc$]  &\\
   \hline
 Ring50  &   50  & 10  & 20  & \textbackslash \\
  \hline
 Fractal3  &   50  & 10  & 20  & 3.0 \\
  \hline
 Fractal2  &   50  & 10  & 20  & 2.0 \\
  \hline
 Fractal1  &   50  & 10  & 20  & 1.0 \\
  \hline
\end{tabular}
\caption{Mock dust distribution parameters adopted.}
\label{tab:models}
\end{table}

To build models Fractal3, Fractal2 and Fractal1 we first added a Gaussian background to Eq.~\eqref{eq:Ring}:
\begin{equation}
    \rho_{RB}(x,y,z) = \rho_{R}(R,z) +\exp\Bigg(\frac{-y^2}{2\sigma_{R}^2}\Bigg)
    \label{eq:ring_background}
\end{equation}
with $\sigma_R=1.0$ kpc that makes it almost constant in the region of interest. Fractal3, Fractal2 and Fractal1 are then obtained by perturbing  Eq.~\eqref{eq:ring_background} with density fluctuations on progressively smaller spatial scales.
They are described by:
\begin{equation}
    \rho_{F}(x,y,z) = \rho_{RB}(x,y,z)\,\textit{F}\,(\textbf{x})
    \label{eq:fractal_ring}
\end{equation}
where $\rho_{RB}(x,y,z)$ is given by Eq. (\ref{eq:ring_background}) and $\bf{x}$$=(x,y,z)$.
$\textit{F}\,(\textbf{x})$ is a fractal field defined as:
\begin{equation}
    F(\textbf{x}) = 1+\big[M(\textbf{x})-1\big]
    \label{eq:fractal_field}
\end{equation}
where $M(\textbf{x})$ is obtained from a 
fixed-amplitude random-phase field
$G(\textbf{x})$ as:
\begin{equation}
    M(\textbf{x})=\rm{exp}\bigg(\textit{G}(\textbf{x})-\frac{\sigma^2}{2}\bigg).
    \label{eq:m_exp_G}
\end{equation}
with $\sigma=1.0$.

To generate $G(\textbf{x})$ we define three wave vectors $k_q$, one for each dimension of the real space. We impose a power spectrum $P(k) \propto k^{-\alpha}$, with $k=|\textbf{k}|=\sqrt{\Sigma_qk_q^2}$ that we adopt to build a field in Fourier space according to:
\begin{equation}
    \tilde{G}(\textbf{k})=\sqrt{P(k)}e^{i\phi(\textbf{k})},
\end{equation}
where $\sqrt{P(k)}$ is the amplitude and $e^{i\phi(\textbf{k})}$ are the phases uniformly distributed in $(0,2\pi)$. We enforce Hermitian symmetry:
\begin{equation}
    \tilde{G}(-\textbf{k})=\tilde{G}(\textbf{k})^*
\end{equation}
so that the resulting field in real space is strictly real. Then, by inverse Fourier transforming $\tilde{G}(\textbf{k})$ we obtain our real Gaussian field $G(\textbf{x})$:
\begin{equation}
    G(\textbf{x}) = \mathcal{F}^{-1}[\tilde{G}(\textbf{k})] .
\end{equation}
Finally, we impose $\langle G\rangle = 0$ and $\rm{Var}(\textit{G})=1$ so that $\langle M\rangle = 1$ and the gas density in models Fractal3, Fractal2 and Fractal1 is quantitatively comparable to that in model Ring50.
Fractal3, Fractal2 and Fractal1 differ for the values of $\alpha$, listed in Table \ref{tab:models}, that determines the spatial scale of the dominant fluctuations: the smaller $\alpha$, the more the small-scale fluctuations dominate.

As a final model we tested a dust distribution from a snapshot of a hydrodynamical simulation presented in \citet{Feng+26} which we refer to as model Hydrosim. The simulation is performed with ISM prescriptions that include multi-phase gas, star formation and stellar feedback with an average spatial resolution of $\sim$2 pc.
\newline

\begin{figure*}
    \includegraphics[width=1\linewidth]{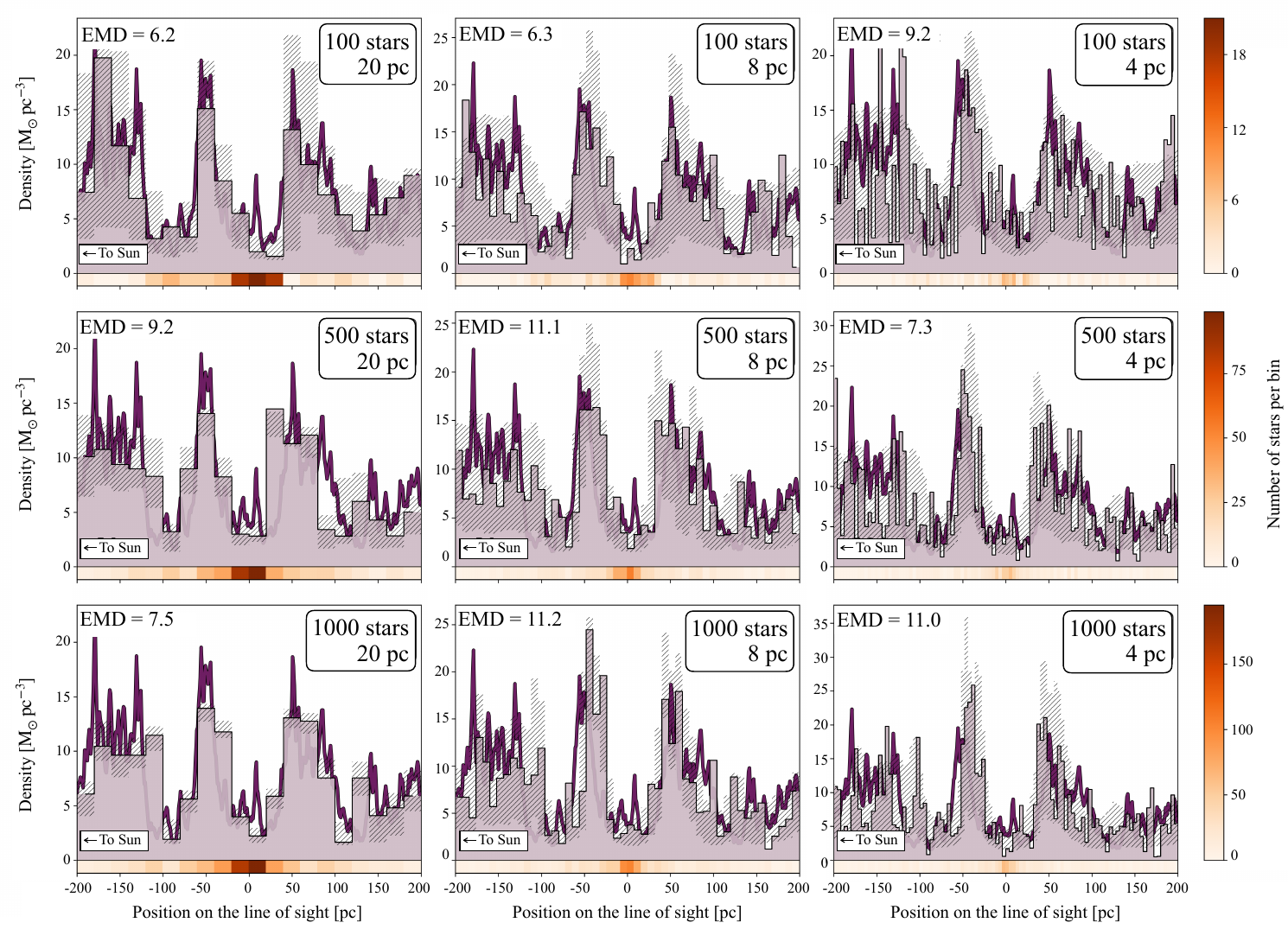}
    \caption{Model Fractal3 reconstructed with different resolutions and star sample sizes. From left to right the adopted resolution is 20 pc, 8 pc and 4 pc. From top to bottom rows increasing star sample sizes as indicated in each panel.}
    \label{fig:1d_fractal3}
\end{figure*}

\subsection{Results}
\label{sec:results}

We present the recovered dust distributions obtained with our method and we show how the results depend on the mock gas density profile (see Table \ref{tab:models}), the number of stars adopted and the spatial resolution along the line of sight.

\subsubsection{Dependence on number of stars and resolution adopted}
\label{sec:star_dependence}

Fig. \ref{fig:megaplot} shows the gas density profile averaged over 500 realisations of the experiment along the representative field centred on ($\ell,b$)=($-5,-5$) pc recovered with our method, compared to the mock gas profile Ring50. The different columns adopt different bin sizes, which represent our spatial resolution, and different rows adopt an increasing number of stars. The true positions of the stars in the samples are represented in the histograms at the bottom of every plot and are binned with the same resolution adopted for the dust profile reconstruction. In the top left corner of every panel we report the median over the 500 realisations of the Earth Mover's Distance (EMD) value that intuitively is the average distance over which the recovered dust density in every bin has to be moved to be reshaped into the mock distribution (see Appendix~\ref{sec:emd} for the details). We use the EMD as a measure of the quality of the reconstructed profiles, namely as the uncertainty in the distance attributed by the method to the dust peaks. Unlike bin-by-bin metrics such as the $\chi^2$, in fact, the EMD is sensitive to the distance over which mass is displaced. Being expressed in parsecs, it can also directly be compared to the resolution adopted. Focusing on the first column where a 20 pc resolution is adopted, two peaks can already be identified with the 100-star sample, but the agreement with the mock profile improves with 200 stars. However, beyond 200 stars, the results do not show improvements regardless of the size of the stellar catalogue adopted because they are limited by the resolution. In fact, we can recover the number and the position of the peaks but neither their intensity nor their width. The EMD values reflect this behaviour.

In the second column we increased the resolution to 8 pc, keeping the same star samples and the results show a general improvement. The 100 star sample provides a weak hint of a double-peaked profile which becomes clear at 200 stars and then improves until 400 stars. Unlike the 20 pc case, at 8 pc the same stellar samples now recover the intensity and width of the peaks as well.

In the third column we tested an even higher resolution --- 4 pc. As expected, 100 stars are not enough to reproduce the mock profile but the agreement improves quickly and saturates at 500 stars. These results clearly show that increasing the resolution can be counterproductive if the catalogue has a very small number of stars or if the features of the profile do not require such a high resolution.

All panels show that on average larger stellar catalogues yield better results. In Fig.~\ref{fig:emd_vs_nstars} we show the median EMD as a function of the number of stars in the catalogue up to 2000 stars. For the 20 pc resolution the trend slightly oscillates around a mean value showing that in this case we are mainly limited by the bin width, whereas the 8 pc and 4 pc curves show a monotonic decrease with increasing size of the catalogue saturating respectively at 500 and 1000 stars. This indicates how many stars we need to reconstruct the mock profile if we require a minimum EMD with a specific confidence level.

\begin{figure}
    \centering
    \includegraphics[width=1.0\linewidth]{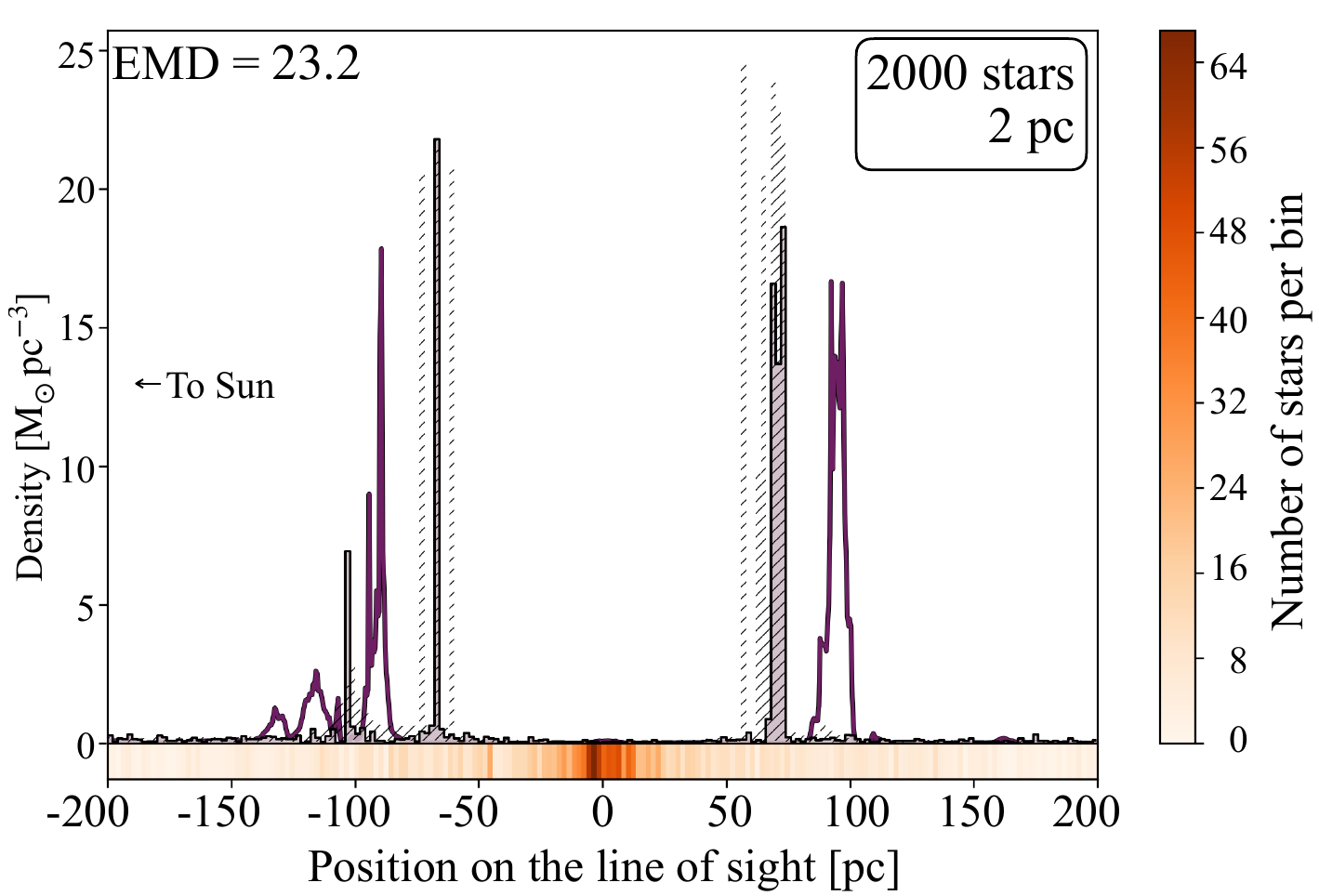}
    \caption{Reconstruction of model Hydrosim with 2 pc resolution and 2000 stars.}
    \label{fig:1d_hydrosim}
\end{figure}

\subsubsection{Dependence on the mock profile}
\label{sec:profile_dependence}
Fig. \ref{fig:1d_fractal3} shows the results for the Fractal3 profile for different star samples and spatial resolutions obtained from a single realisation of the experiment for each combination. Unlike Ring50, the features of Fractal3 require higher resolutions to be recovered.
Models Fractal2 and Fractal1 show the same behaviour as Fractal3 but are presented in Appendix \ref{sec:app_fractals}.

Fig. \ref{fig:1d_hydrosim} shows the result for the Hydrosim model obtained with 2000 stars and 2 pc resolution. We can recover the number of peaks, whether they are on the near or far side and their intensity. However, the peaks show a systematic offset towards the centre. This can be attributed to the relatively large $\sigma_j$ (and corresponding wide $P(\Sigmaobs|\Sigma)$) adopted (see Eq.~\ref{eq:p_sigma}) that results in the dust peaks to move more towardsthe centre where $P(y)$ grows, keeping $P(\Sigmaobs|\Sigma)$ almost constant. A smaller $\sigma_j$ (i.e. narrower $P(\Sigmaobs|\Sigma)$) would constrain better the position of the peaks but it would also generate numerical instabilities when dealing with the stars in the near side of the line of sight, thus worsening the foreground of the recovered profile.

Figs. \ref{fig:megaplot} - \ref{fig:1d_hydrosim} show that small samples can capture the overall shape of the profile and distinguish foreground from background clouds thanks to the NSD stellar distribution peaking at the Galactic centre. On the other hand, larger star samples allow higher-resolution grids producing more detailed dust profiles. 

\subsubsection{2D and 3D maps}
\label{sec:2d_3d_maps}

In our framework we treat each line of sight independently. Therefore to obtain the 2D and 3D maps we apply the method to neighbouring lines of sight and stack the 1D results.

\begin{figure}
    
    \includegraphics[width=1.0\linewidth]{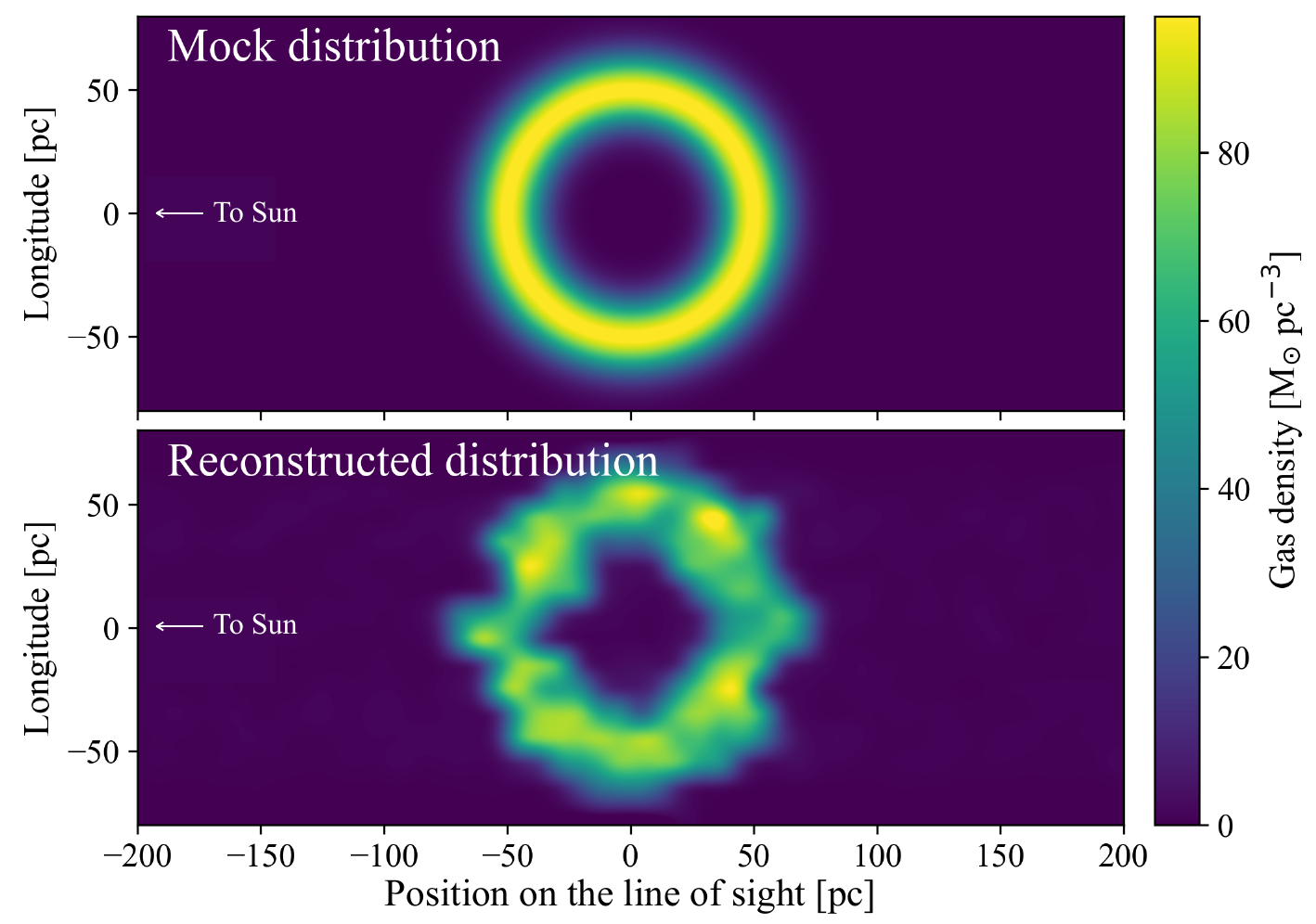}
    \caption{Comparison between 2D map of the model Ring50 on the longitude-line of sight plane (top panel) and the map recovered with the presented method (bottom panel). 1000 stars for each line of sight and a 4 pc resolution were adopted. The x axis is the same as Figs. \ref{fig:megaplot}-\ref{fig:1d_fractal3} and y axis is the longitude spanned by the gas distribution. The direction pointing towards the Sun is also indicated.}
    \label{fig:2d_ring}
\end{figure}

Fig. \ref{fig:2d_ring} shows the 2D map of the plane $z=-5$ pc obtained with our method compared to Ring50. The general shape and intensity are recovered with small localised distortions affecting some regions of the map. 

\begin{figure}
    \centering
    \includegraphics[width=1.0\linewidth]{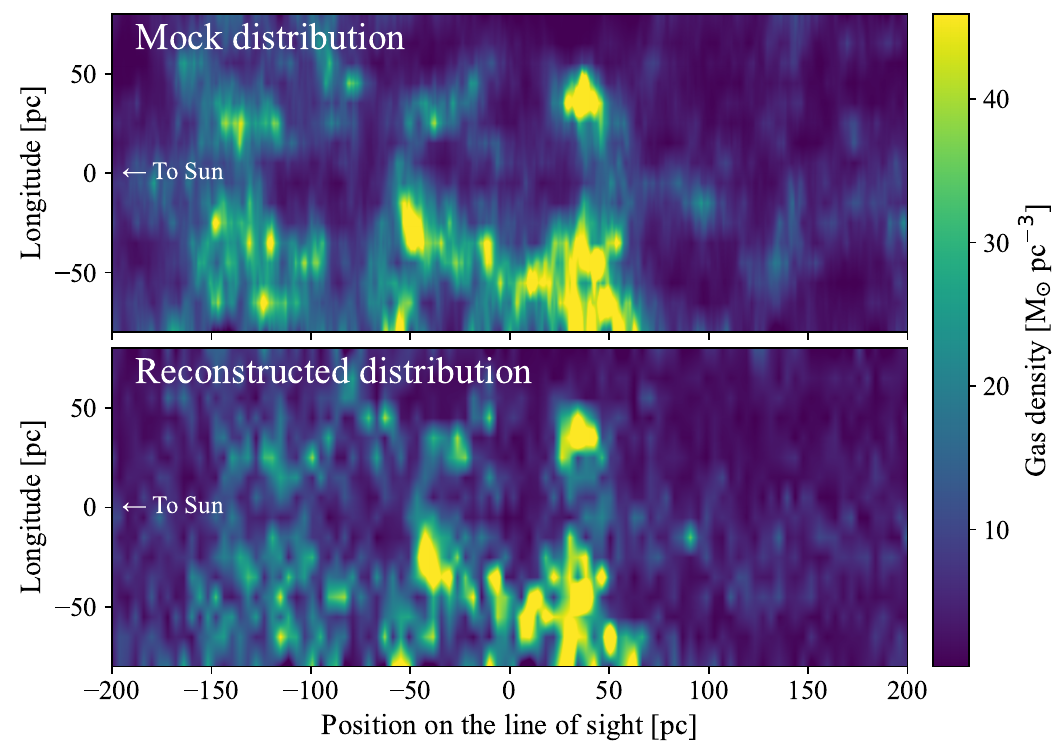}
    \caption{Same as Fig. \ref{fig:2d_ring} for model Fractal3.}
    \label{fig:2d_fractal3}
\end{figure}

\begin{figure*}
    \centering
    \includegraphics[width=1.0\linewidth]{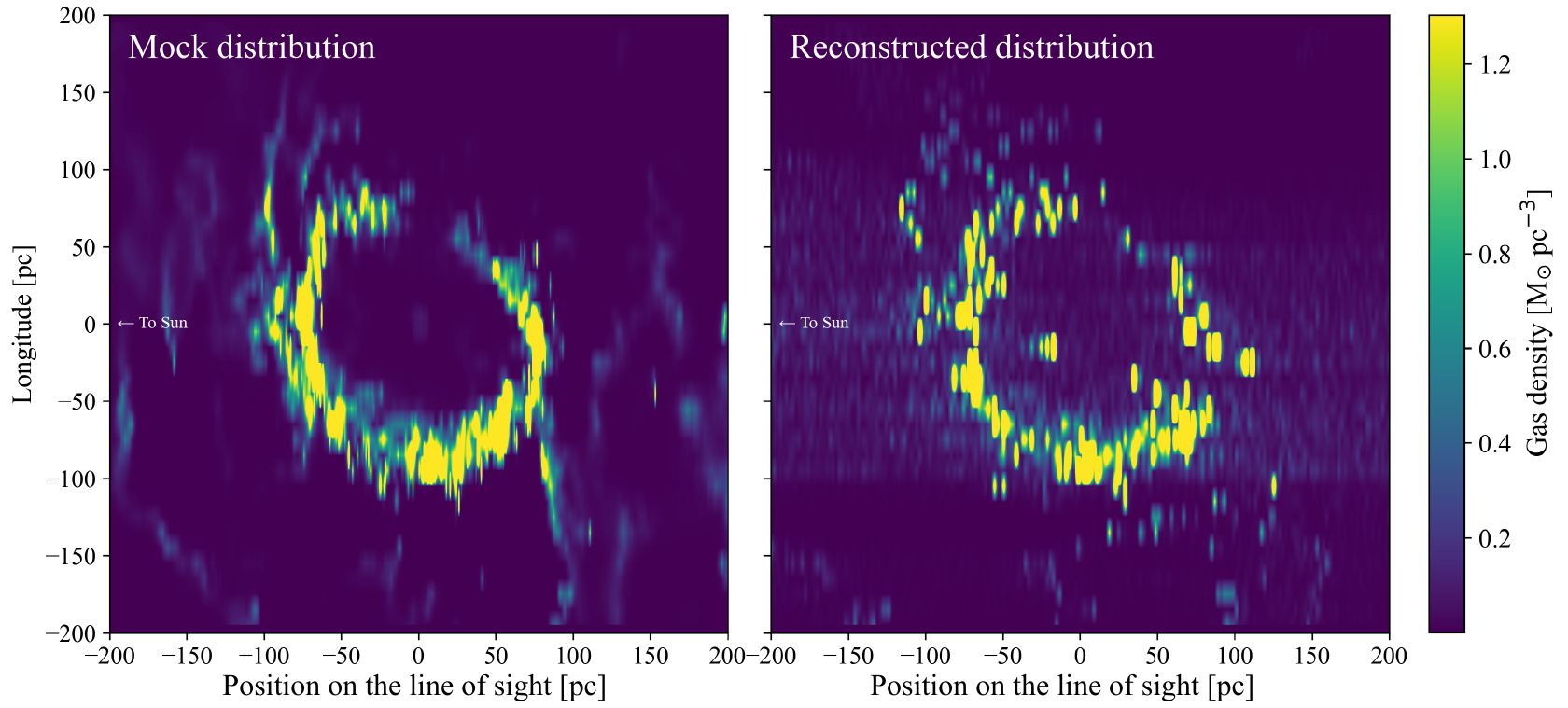}
    \caption{Same as Fig. \ref{fig:2d_ring} for model Hydrosim.}
    \label{fig:2D_hydrosim}
\end{figure*}

The 2D results for the model Fractal3 and Hydrosim, shown in Figs. \ref{fig:2d_fractal3} and \ref{fig:2D_hydrosim} respectively, also show good agreement between the mock profiles and the reconstructed ones.
We show the 2D maps of Fractal2 and Fractal1 in Appendix \ref{sec:app_fractals}.

In Figs. \ref{fig:3d_ring}, \ref{fig:3d_fractal3} and \ref{fig:3d_hydrosim} we show the 3D maps of the Ring50, Fractal3 and Hydrosim models, respectively. They all faithfully reproduce the overall mock distributions both in the shape and in the intensity of the clouds. However, the reconstruction of Ring50 shows some artificial clumpiness that was not present in the mock distribution. The 3D maps of Fractal2 and Fractal1 are shown in Appendix \ref{sec:app_fractals}.

\begin{figure}
    \centering
    \includegraphics[width=1.0\linewidth]{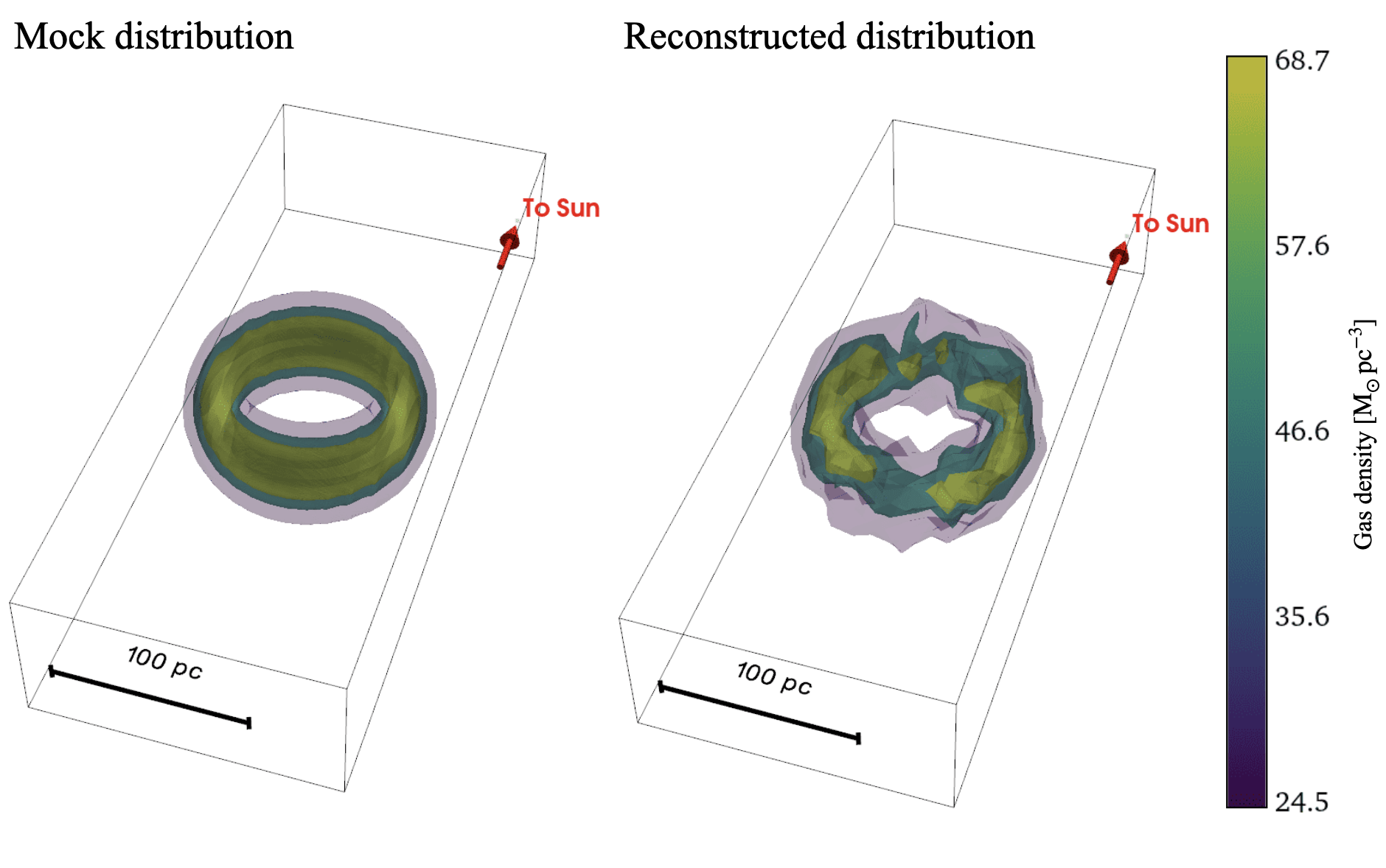}
    \caption{Comparison between the 3D map of the mock model Ring50 on the left with the recovered distribution on the right. Four different isodensity surfaces are plotted colour coded with the gas density. The direction towards the Sun is indicated by the red arrow.}
    \label{fig:3d_ring}
\end{figure}

\begin{figure}
    \centering
    \includegraphics[width=1.0\linewidth]{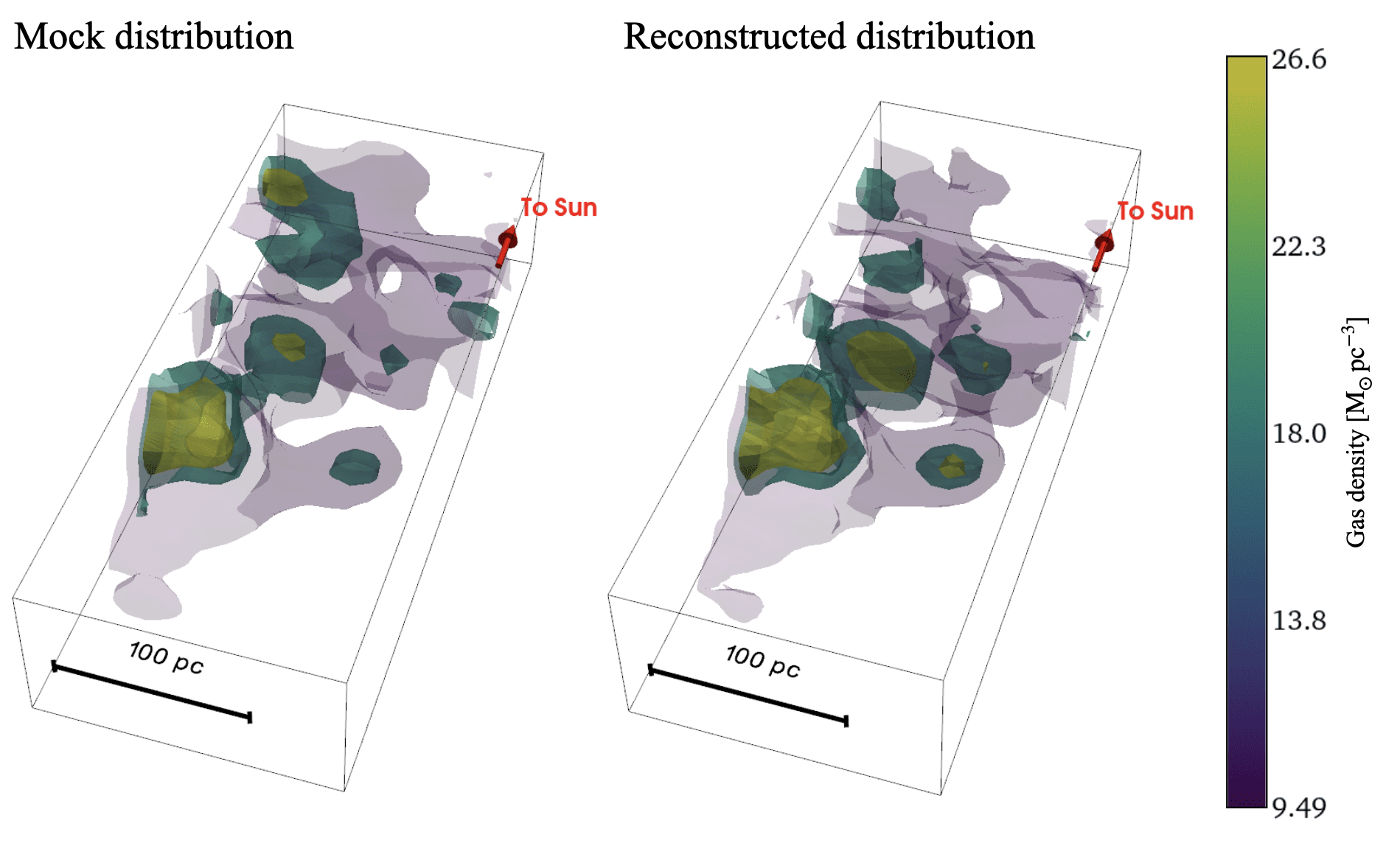}
    \caption{Same as Fig. \ref{fig:3d_ring} for model Fractal3.}
    \label{fig:3d_fractal3}
\end{figure}

\begin{figure*}
    \centering
    \includegraphics[width=1.0\linewidth]{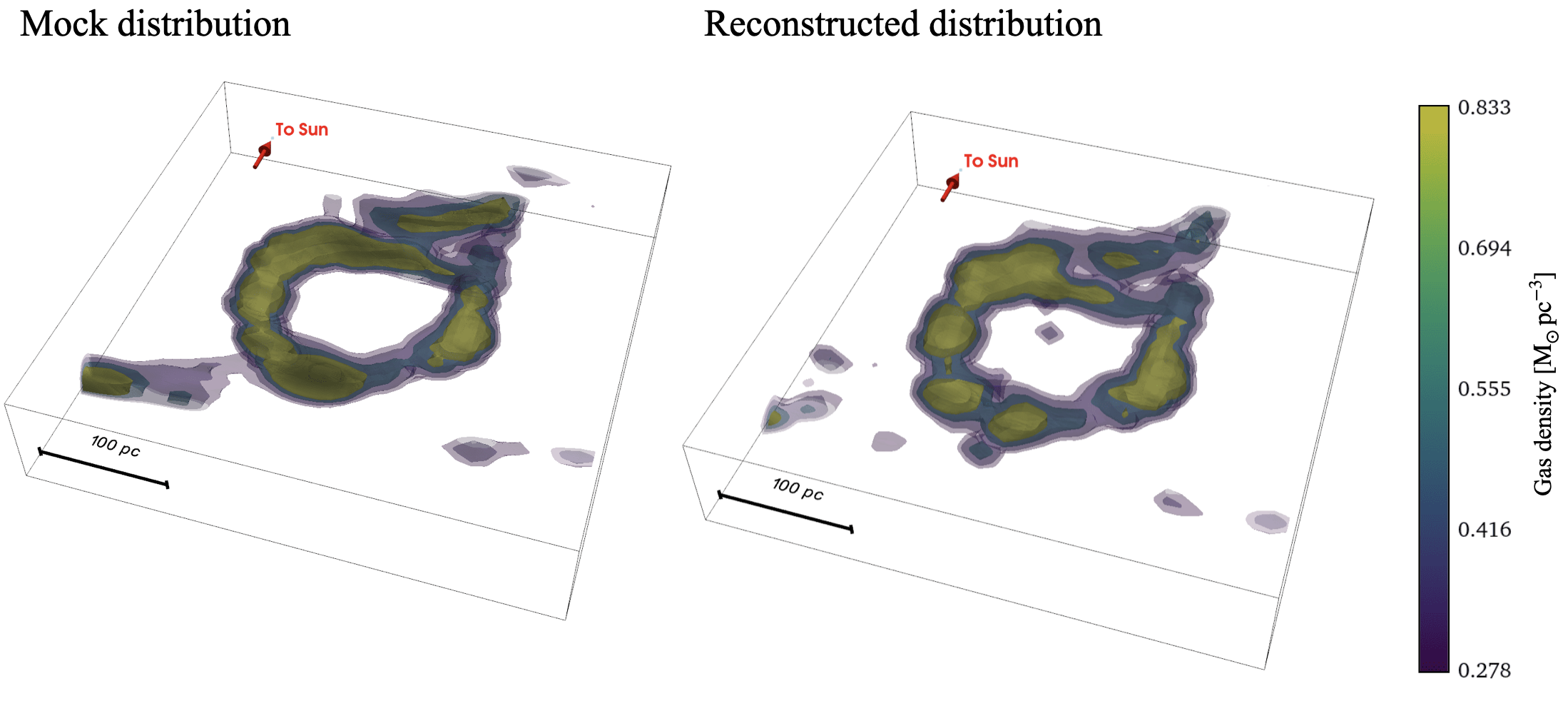}
    \caption{Same as Fig. \ref{fig:3d_ring} for the model Hydrosim.}
    \label{fig:3d_hydrosim}
\end{figure*}

\section{Discussion}
\label{sec:discussion}

\subsection{Spatial resolution limit along the line of sight}
\label{sec:spatial_resolution}

\begin{figure}
    \centering
    \includegraphics[width=1.0\linewidth]{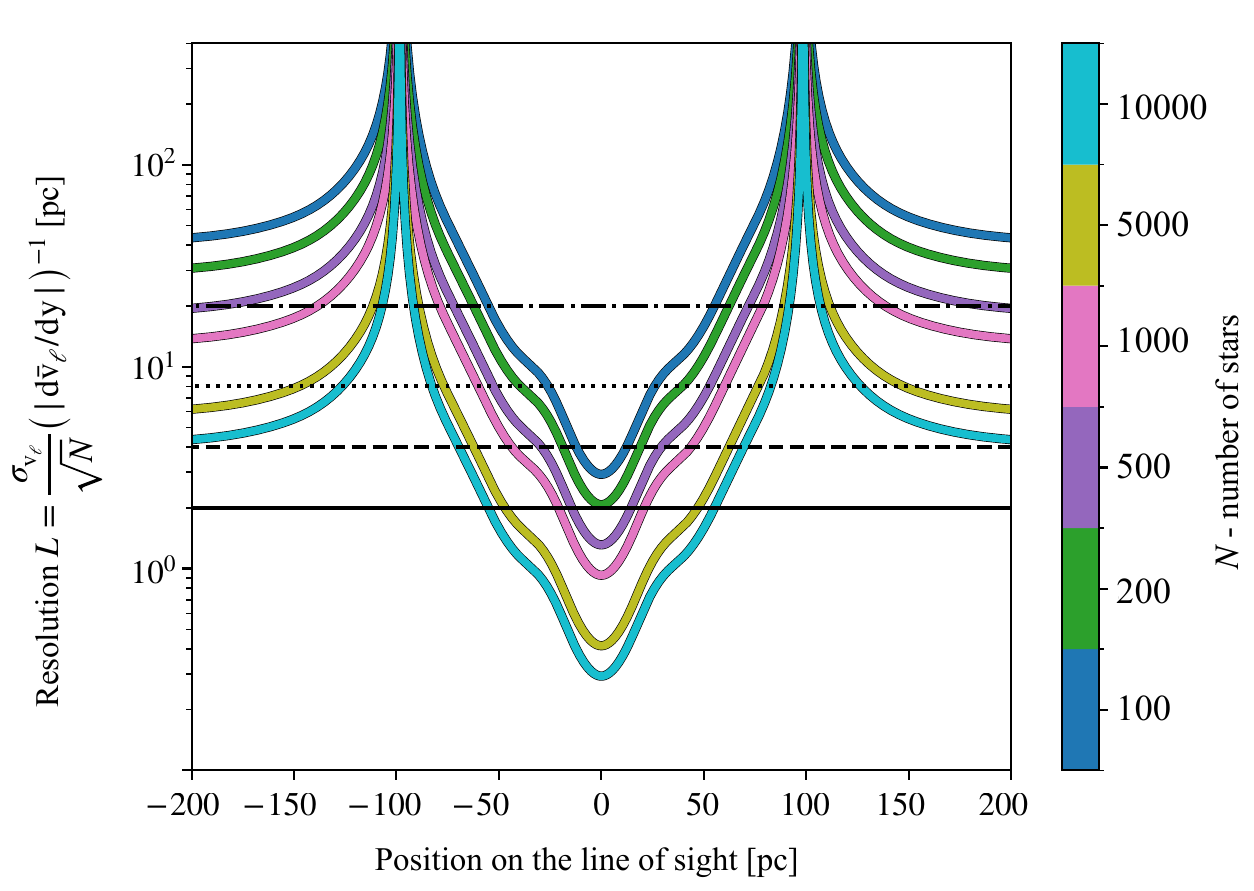}
    \caption{Dynamical resolution along the line of sight ($-5$,$-5$)~pc for different numbers of stars adopted. The dotted-dashed, dotted, dashed and solid lines mark the four resolution adopted in our tests.}
    \label{fig:resolution}
\end{figure}

The spatial resolution of the recovered dust profile is set by how quickly the rotation velocity changes along the line of sight (see Fig.~\ref{fig:proper_motion}), compared to the uncertainty on that velocity. We estimate the limit of the spatial resolution as:
\begin{equation}
    L(\ell,y,b)=\frac{\sigma_{v_\ell}(\ell,y,b)}{\Big|\frac{\rm{d} \bar{\textit{v}}_\ell}{\rm{d}\textit{y}}\Big|\sqrt{N}}
    \label{eq:resolution}
\end{equation}
where $\rm{d} \bar{\textit{v}}_\ell / \rm{d}\textit{y}$ is the rotation velocity gradient and $\sigma_{v_{\ell}}(\ell,y,b)$ is the dispersion on the rotation velocity, both provided by the dynamical model, and $N$ is the number of stars in the catalogue. Consider two stars at positions $y_1$ and $y_2$: if the gradient is steep compared to $\sigma_{v_{\ell}}$, even a small separation $|y_1-y_2|$ produces distinguishable velocities, and the likelihood can resolve the dust density between the two positions. If instead the gradient is shallow compared to $\sigma_{v_{\ell}}$, the velocities of the two stars overlap unless $|y_1-y_2|$ is large enough to compensate for the shallow gradient.
An additional factor contributing to the resolution is the size of the stellar catalogue: larger catalogues provide a finer sampling of the column density on the line of sight.
Fig. \ref{fig:resolution} shows $L(y)$ for different $N$ and a comparison with the four resolutions adopted in the tests presented above. The resolution varies along the line of sight according to how rapidly the velocity changes. The best resolution is found in the GC where $\bar{v}_\ell$ goes from positive to negative values and the worst is located at $-$100 pc and 100 pc, where the $\bar{v}_\ell$ has a maximum and a minimum respectively. The divergence of $L(y)$ in $-$100 pc and 100 pc is an artefact of our implementation and depends on the fact that the potential adopted in the dynamical model includes the NSD only. The addition of other components, such as the Galactic bar, would remove the feature.
Fig. \ref{fig:resolution} shows that the quality of our recovered dust profiles can still be improved in the central regions of the line of sight by adopting non-uniform bins.

\subsection{Proper motions as a safeguard against selection effects}
\label{sec:no_proper_motions}
Throughout the paper we have stressed the central role of the proper motions in our method. However, nothing in the formal derivation of Eq.~\eqref{eq:likelihood_sigma} strictly requires them. Following the derivation in the absence of proper motion measurements we arrive at:
\begin{equation}
    \mathcal{L}_j(\Sigmaobs[j]|\Sigma) = \int P(\Sigmaobs[j]|\Sigma,y) P(y)
     \, \di y \,.
\label{eq:like_no_v2}  
\end{equation}
Does this likelihood in principle allow us to recover the dust profile? Does it work in practice? To answer these questions, we performed the following two experiments.

\begin{figure*}
    \centering
    \includegraphics[width=1.0\linewidth]{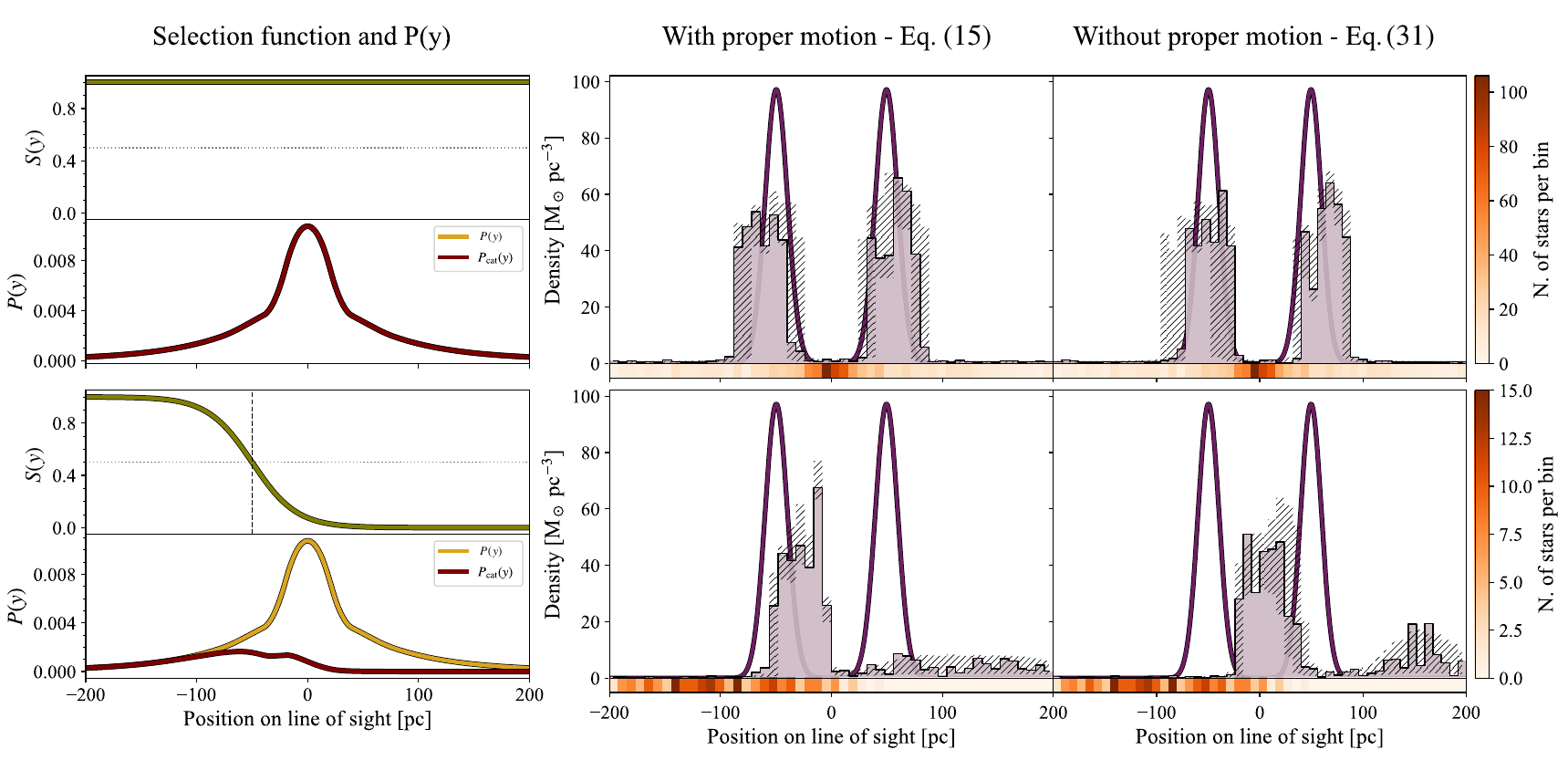}
    \caption{Effect of removing the proper motion information. All panels show the reconstruction of Ring50 along the line of sight $(\ell,b)=(-5,-5)$~pc with 8~pc resolution, using 1000 mock stars. In the left panels, $P_{\rm cat}(y)=S(y)\,P(y)$ (red line) denotes the distribution of stars in the mock catalogue along the line of sight, while $P(y)$ (yellow line) is the distribution of stars along the light of sight assumed in the likelihood. In the top row we assume that $P_{\rm cat}(y)=P(y)$, which is equivalent to perfectly modelling both the selection function and the stellar distribution along the line of sight. In the bottom row, $P_{\rm cat}(y) \neq P(y)$, which mimics the situation in which the selection function is modelled incorrectly. The right panels show the profiles recovered with (Eq.~\ref{eq:likelihood_sigma}) and without (Eq.~\ref{eq:like_no_v2}) proper motions. When the selection function is perfectly known (top row), the two reconstructions are of comparable quality. When the selection function is modelled incorrectly (bottom row), the near-side peak is wrongly placed at the centre of the line of sight in the absence of proper motion information, while with proper motions it is moved back close to its correct position.}
    \label{fig:ginsburg}
\end{figure*}

In the first experiment, we reconstructed the Ring50 profile along the same line of sight as in Fig.~\ref{fig:megaplot}, adopting 1000 stars and a resolution of 8~pc, using both Eq.~\eqref{eq:likelihood_sigma} (that includes the proper motion information) and Eq.~\eqref{eq:like_no_v2} (that does not). The results are shown in the top row of Fig.~\ref{fig:ginsburg}. Surprisingly, the two recovered profiles are very similar: dropping the proper motion term does not significantly degrade the performance of the method. One might therefore conclude that the proper motions are unnecessary. However, this conclusion would be misleading.

The reason the method works without proper motions in this first experiment is that the stellar distribution assumed in the likelihood, $P(y)$ in Eq.~\eqref{eq:P_y} (yellow line in Fig.~\ref{fig:ginsburg}), is identical to the distribution from which the stars in the catalogue are actually drawn $P_{\rm cat}$ (red line in Fig.~\ref{fig:ginsburg}). This is equivalent to assuming that we know perfectly both the stellar density along the line of sight and the selection function. Since $\Sigma(y)$ is a cumulative quantity that never decreases along the line of sight, the stars with the lowest $\Sigmaobs[j]$ are, statistically, the nearest ones. If $P_{\rm cat}(y)$ is known exactly, ranking the stars by extinction and matching this ranking to the cumulative distribution of $P_{\rm cat}(y)$ is sufficient to place the stars, and hence the dust, at approximately the correct distances.

With real data, however, the stellar distribution of the catalogue $P_{\rm cat}(y) = S(y|\Sigma) P(y)$, which is the product of the selection function of the survey and the stellar density distribution along the line of sight, is only known approximately. In the second experiment (bottom row of Fig.~\ref{fig:ginsburg}) we mimic this situation. We assume that the probability of detecting a star decreases with distance according to a selection function $S(y|\Sigma) = S(y)$ that depends only on the distance (green line in the bottom-left panel of Fig.~\ref{fig:ginsburg}), which drops from unity to zero around $y=-50$~pc (dashed vertical line). This selection function is chosen for illustrative purposes, but it qualitatively mimics the loss of the stars behind the CMZ due to extinction. As a result, the stars in the mock catalogue follow the distribution $P_{\rm cat}(y)$ (red line), with almost no stars on the far side. We then pretend that we do not know the selection function and reconstruct the dust profile with both likelihoods assuming a (wrong) flat selection function. In other words, the likelihood assumes that the stars are distributed according to $P(y)$ (yellow line), whereas they actually follow $P_{\rm cat}(y)$.

Without proper motions (Eq.~\ref{eq:like_no_v2}, bottom-right panel of Fig.~\ref{fig:ginsburg}), the near-side peak is incorrectly placed at the centre of the line of sight, where there is no dust. This happens because, owing to the selection function, a much larger fraction of the catalogue stars lies in front of the near-side cloud than the likelihood expects. Under the assumed $P(y)$, the same fraction of stars is reached only close to $y\simeq0$, so the jump in extinction, and hence the cloud, is placed there. When the proper motions are taken into account (Eq.~\ref{eq:likelihood_sigma}, bottom-central panel), the peak moves back towards its correct position on the near side. This is because the proper motions carry direct information on the distance of each star, independently of the assumed $P(y)$, and can therefore correct the placement dictated by the wrong stellar distribution. The far-side peak is not recovered with either likelihood, because almost no stars are present on the far side, so the data contain no information on the dust there. The corresponding column density, which must still be assigned somewhere because $\Sigmatot$ is fixed (Eq.~\ref{eq:softmax}), is spread out over the far side of the line of sight. 

In conclusion, the proper motions make the method robust against errors in the modelling of the selection function and/or of the line-of-sight distribution of stars. This is important because, given the high density and the clumpy and filamentary structure of the interstellar medium in the GC, the selection function of real surveys will be complex, difficult to model, and can differ strongly between lines of sight. Including the proper motions therefore provides a safeguard against errors in the characterisation of the selection function, although, like any method based on stellar extinctions, it cannot constrain the dust in regions that contain no stars.

\section{Conclusions and future prospects}
\label{sec:conclusion}
We developed a new method to determine the three-dimensional structure of the CMZ by exploiting the observations of the stellar proper motions and extinctions. With this proof of concept we demonstrated the validity and robustness of the method by testing its performance on mock dust distributions and mock stars in preparation for its application to real data. 
We conclude that:
\begin{enumerate}
    \item small star samples ($\sim$ 100 stars per line of sight) are effective at highlighting the larger scale features of the dust profiles (20 pc resolution);
    \label{item_1}
    \item larger star samples ($\gtrsim$ 1000 stars per line of sight) are useful to recover the smaller dust features at higher resolution (8,4 and 2 pc);
    \label{item_2}
    \item when the selection function is well known the method works even in the absence of proper motions. When the selection function is not or only approximately known, the inclusion of the proper motions can make the method more robust so that it can still provide meaningful results;
    \item the spatial resolution limit of our method is set by the ratio of the stellar velocity dispersion to the gradient of the rotation velocity as a function of position of the line of sight.
\end{enumerate}
 
The key advantage of our method is that it is completely non-parametric and avoids any a priori assumption on the shape of the gas profile. The assumptions on our parameters $\Theta$ (see Sect. \ref{sec:method}) are in fact solely introduced to preserve their physical meaning, namely by fixing their sum to the observed value and enforcing non-negativity. As a result, all the profiles presented in Sect. \ref{sec:results} emerge directly from the mock data used. 
Moreover, our method is not only non-parametric along individual lines of sight but also across the full two- and three-dimensional space. Although the lines of sight are treated independently, the coherence between neighbouring directions arises naturally in the reconstructed profiles, supporting the robustness of the method presented.

In conclusion, the mock tests show that our method has the potential to be a powerful tool to reconstruct the true 3D geometry of the CMZ when applied to real data such as those by the KMOS survey \citep{Fritz+21}, the GALACTICNUCLEUS survey \citep{Nogueras+19,Shahzamanian+19,Shahzamanian+22}, the VVV survey \citep{Minniti+10_VVV}, the JWST/NIRCam Legacy Survey of the Galactic Center (GO 10678; see \citealt{Schoedel+23}) and the NSC observations by \citet{Hosek+25}. 
The $\sim 100$-$1000$ stars per beam that we require to recover the dust profiles correspond to $\sim 1$-$10$ stars $\rm{pc}^{-2}$, once we consider that the beams adopted have cross section 10 pc $\times$ 10 pc. These requirements are within reach for the above-mentioned surveys: GALACTICNUCLEUS combined with the HST proper motions provides $\sim 25$ stars $\rm{pc}^{-2}$, VIRAC2 \citep{Smith+25_VIRAC2} built on the VVV data provides $\sim 50$ stars $\rm{pc}^{-2}$ in the GC and JWST/NIRCam Legacy Survey is expected to collect proper motions for $\sim 150$ stars $\rm{pc}^{-2}$ after one epoch and $\sim 1500$ stars $\rm{pc}^{-2}$ after two epochs.

Additional upcoming surveys such as the KMOS VVVX-GalCen\footnote{https://www.eso.org/sci/observing/PublicSurveys/KMOS-surveys-projects.html} Spectroscopic Survey \citep{Arnaboldi+26}, the Gas and dust in and in front of the Galactic Center\footnote{https://roman.ipac.caltech.edu/cycle1-approved-programs/19008}, the Galactic Bulge Time Domain Survey and the Roman Galactic Plane Survey \citep{RGPS} all planned with the Nancy Grace Roman Space Telescope \citep{Terry+23}, the MOONS Surveys \citep{Gonzalez+20_MOONS}, the JASMINE mission \citep{Kawata+24} and the JWST NSC survey\footnote{https://www.stsci.edu/jwst/phase2-public/5368.pdf} (Do et al. in prep.) will largely extend the already available datasets in the near future.

However, before applying our method to real datasets, additional phenomena must be taken into account and included. We plan to:
\begin{itemize}

\item include other Milky Way components in our framework, such as the NSC and the Galactic bar;

\item include the uncertainties on the velocity $v_{\ell}$, or equivalently on $\mul$, as presented in Appendix \ref{sec:error_vl};

\item include also $v_{\rm{los}}$ and $\mu_b$ where available to increase accuracy

\item relax the assumption $P(y|\Sigma)=P(y)$ by including the modelling of the selection function as anticipated in Sect. \ref{sec:P_y}.
\end{itemize}

\begin{acknowledgements}
    
We thank the anonymous referee for the constructive comments and suggestions that significantly improved the paper.
AV, MCS, MD, ZF, KF, and XL acknowledge financial support from the European Research Council under the ERC Starting Grant “GalFlow” (grant 101116226). AV, MCS, MD, ZF, KF, and XL also acknowledge support from CINECA/ISCRA, which awarded the class B project “GCflow” access to the LEONARDO supercomputer, owned by the EuroHPC Joint Undertaking, hosted by CINECA (Italy). MCS further acknowledges financial support from the Fondazione Cariplo under the grant ERC attrattivit\`{a} n. 2023-3014.
FNL  and RS acknowledge financial support from the Severo Ochoa grant CEX2021-001131-S funded by MCIN/AEI/ 10.13039/501100011033 and from grants PID2022-136640NB-C21 and PID2024-162148NA-I00 funded by MCIN/AEI 10.13039/501100011033 and by the European Union. FNL acknowledges financial support from the Ramón y Cajal programme (RYC2023-044924-I) funded by MCIN/AEI/10.13039/501100011033 and FSE+. CB  gratefully  acknowledges  funding  from  National  Science  Foundation  under  Award  Nos. 2108938, 2206510, 2414862, and CAREER 2145689, as well as from the National Aeronautics and Space Administration through the Astrophysics Data Analysis Program under Award ``3-D MC: Mapping Circumnuclear Molecular Clouds from X-ray to Radio,” Grant No. 80NSSC22K1125 as well as participation in the PRIMA project under Grant No. 80NSSC25K7944. DL gratefully acknowledges funding from the National Science Foundation under Award Nos. 1816715, 2108938, and CAREER 2145689; and NASA FINESST Award No: 80NSSC24K1474. We thank John Magorrian and Eugene Vasiliev for the useful discussions and the valuable suggestions. 
\end{acknowledgements}

\bibliography{method}

\begin{appendix}

\section{Implementation of the uncertainty on $v_{\ell}$ in the Likelihood}
\label{sec:error_vl}

Let $v_{\ell,j}$ be the true velocity of the $j$-th star, $v_{\ell,j,\rm{obs}}$ the observed velocity and $\delta v_{\ell,j}$ the observational error. In this more general case Eq. \eqref{eq:likelihood_v} becomes:
\begin{equation}
    \mathcal{L}_j (v_{\ell,j,\rm{obs}},A_j | \Sigma) = \int P(v_{\ell,j,\rm{obs}}|y) P(A_j|\Sigma,y) P(y|\Sigma) \di y \,,
    \label{eq:likelihood_v_error}
\end{equation}
where
\begin{equation}
    P(v_{\ell,j,\rm{obs}}|y)=\int P(v_{\ell,j,\rm{obs}}|\,v_{\ell,j})P(v_{\ell,j}|y)\di v_{\ell}\,,
    \label{eq:v_error}
\end{equation}
where $P(v_{\ell,j}|y)$, as before, is given by the dynamical model of choice and $P(v_{\ell,j,\rm{obs}}|\,v_{\ell,j})$ can be parameterised as a Gaussian distribution of width $\delta v_{\ell,j}$ as:
\begin{equation}
    P(v_{\ell,j,\rm{obs}}|\,v_{\ell,j}) = \frac{1}{\sqrt{2\pi}\delta v_{\ell,j}}\rm{exp}\Bigg(\frac{-(\textit{v}_{\ell,j,\rm{obs}} - \textit{v}_{\ell,j})^2}{2\delta \textit{v}_{\ell,j}^2}\Bigg)\,.
    \label{eq:v_error_gaussian}
\end{equation}
When assuming $\delta v_{\ell,j}=0$, as we do in this paper, $v_{\ell,j,\rm obs}=v_{\ell,j}$, Eq.~\eqref{eq:v_error_gaussian} collapses to a Dirac delta and Eq.~\eqref{eq:likelihood_v_error} reduces to Eq.~\eqref{eq:likelihood_v}.

\section{Earth Mover's Distance}
\label{sec:emd}

The Earth Mover's Distance (EMD) \citep{Rubner+2000,Villani+2009} originates from an optimisaion problem: finding the transport plan to reshape one distribution into another one at the minimum cost by moving mass. The cost is defined by taking the mass moved and weighting it by the distance over which it is moved and then integrating over all the mass. The EMD is the value of this minimum cost.

In 1D the EMD is defined as follows. Let $\rho(y)$ and $\tilde{\rho}(y)$ be two 1D distributions defined in the interval $[y_{\rm{min}},y_{\rm{max}}]$ with the same total mass $M_{\rm{tot}}$, and the cumulative distributions be respectively $F(y)=1/M_{\rm{tot}} \int_{y_{\rm{min}}}^y \rho(y')\di y'$ and $\tilde{F}(y) =1/M_{\rm{tot}} \int_{y_{\rm{min}}}^y \tilde{\rho}(y')\di y'$, then the EMD \citep{Rubner+2000,Villani+2009} between the two is defined as:
\begin{equation}
    \mathrm{EMD} = \int_{y_{\rm{min}}}^{y_{\rm{max}}} \left| F(y) - \tilde{F}(y) \right| \, \mathrm{d}y \, .
    \label{eq:emd}
\end{equation}
Geometrically, the EMD is the area enclosed between the two cumulative curves. Eq.~\eqref{eq:emd} is symmetric, it vanishes only for $\tilde{\rho}(y)=\rho(y)$ and it satisfies the triangle inequality, which ensures that the EMD is a metric.

In our case, Eq.~\eqref{eq:emd} translates into:
\begin{equation}
    \mathrm{EMD} = \Delta y\sum_{k=1}^{M-1}\Big|F_k - \tilde{F}_k \Big|
    \label{eq:emd_mycase}
\end{equation}
expressed in pc, that has been adopted to compute the EMD throughout the paper.
Here, $\Delta y$ is the bin size adopted, and $F_k$ and $\tilde{F}_k$ are the normalised cumulative column densities of the mock and recovered profiles defined as $F_k= 1/\Sigmatot \big(\sum_{i\leq k}\Sigma_i^{\rm{mock}}\big)$ and $\tilde{F}_k= 1/\Sigmatot \big(\sum_{i\leq k}\Sigma_i\big)$. $\Sigma_i^{\rm{mock}}$ is the mock profile integrated over the same bins as the recovered one, so that the EMD is not affected by structures on scales smaller than the resolution imposed by construction.

\section{Models Fractal2 and Fractal1}
\label{sec:app_fractals}

\begin{figure*}
    \centering
    \includegraphics[width=1\linewidth]{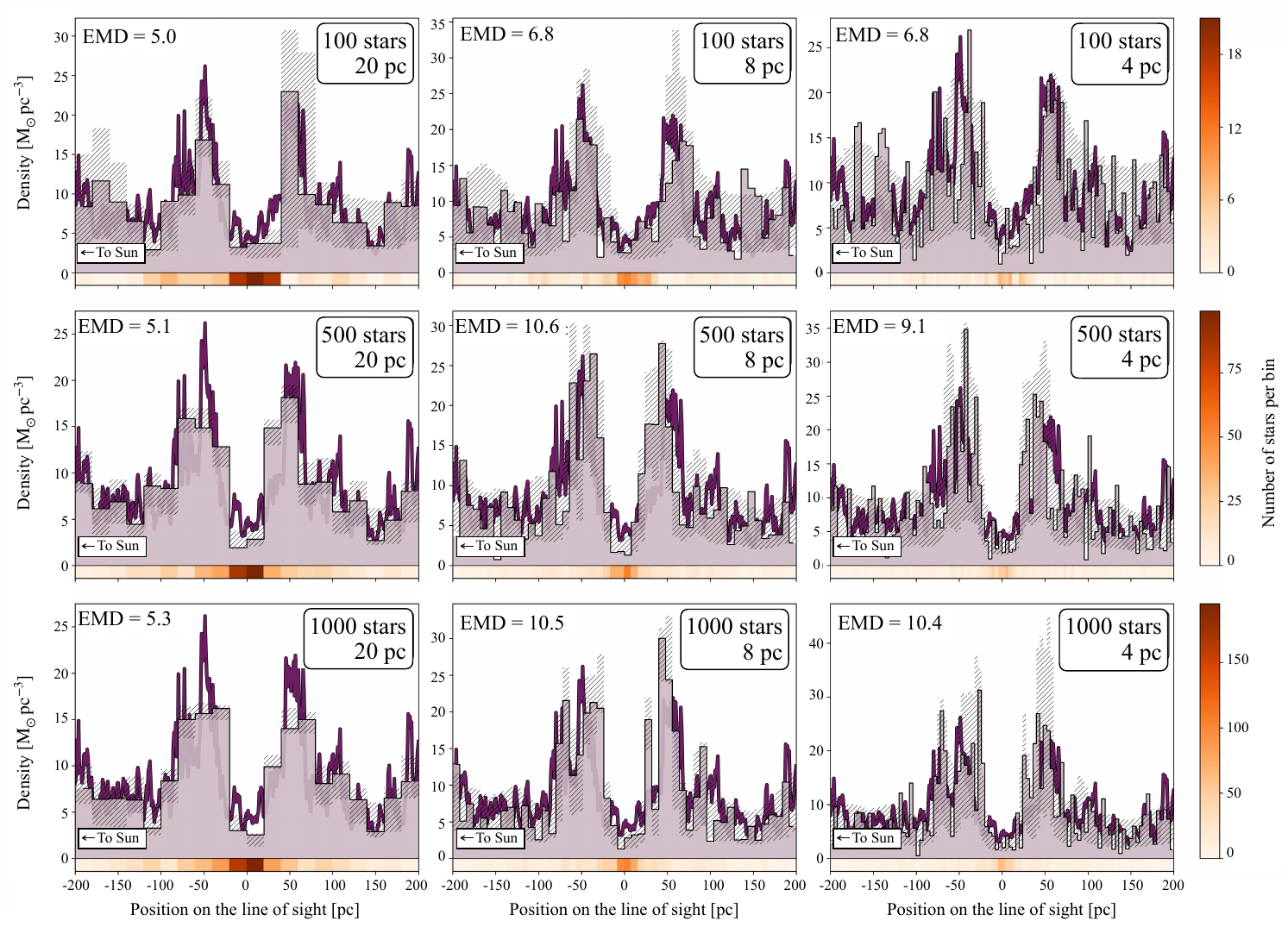}
    \caption{Same as Fig. \ref{fig:1d_fractal3} for model Fractal2.}
    \label{fig:1d_fractal2}
\end{figure*}

\begin{figure*}
    \centering
    \includegraphics[width=1\linewidth]{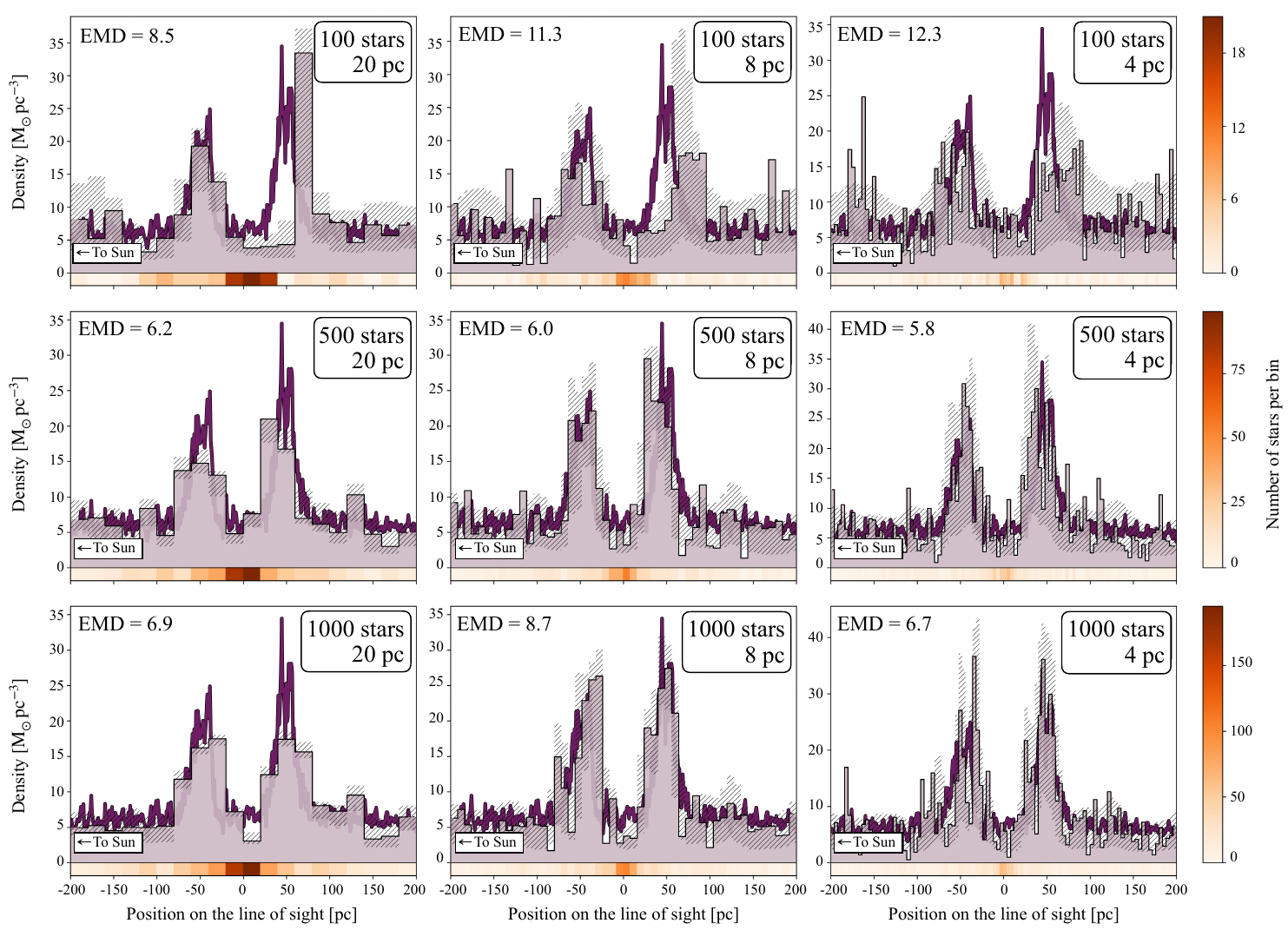}
    \caption{Same as Fig. \ref{fig:1d_fractal3} for model Fractal1.}
    \label{fig:1d_fractal1}
\end{figure*}

Models Fractal2 and Fractal1 are shown in Figs. \ref{fig:1d_fractal2} and \ref{fig:1d_fractal1}, respectively. The results in these two cases show the same behaviour as Fractal3 (see Sect. \ref{sec:profile_dependence}) in terms of dependency on resolution and star samples. By using our method we are still able to recover the trend of the mock gas profiles and with the high resolution set up the method is effective at identifying the smaller peaks as well. 

\begin{figure}
    \centering
    \includegraphics[width=1.0\linewidth]{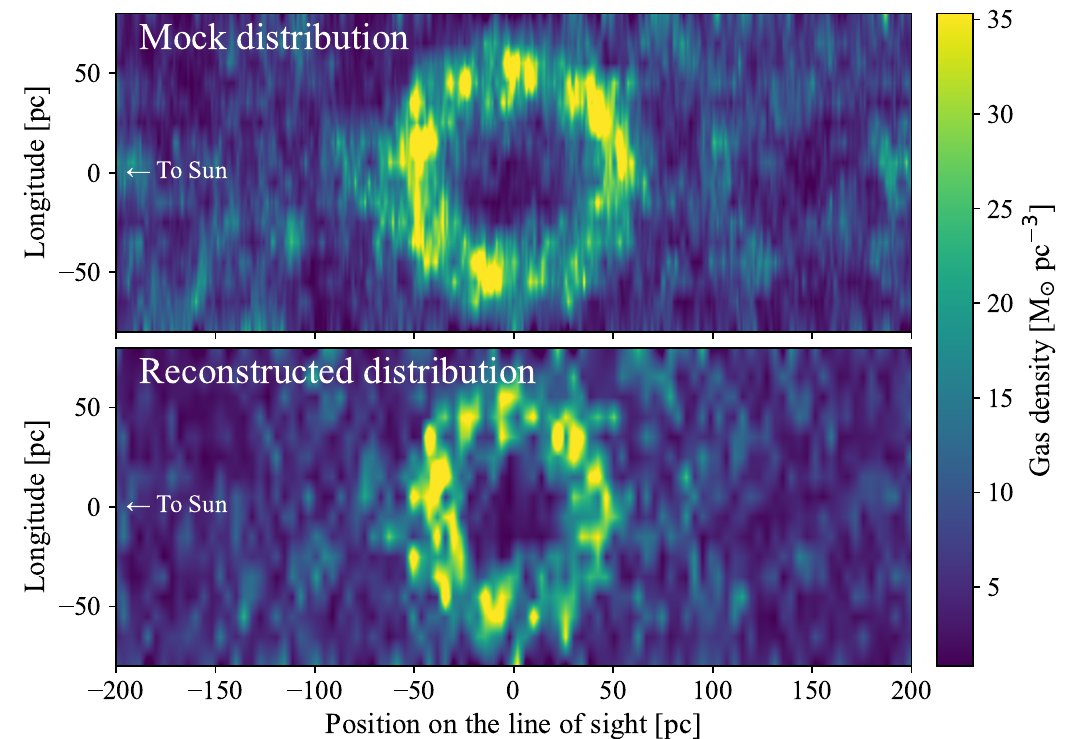}
    \caption{Same as Fig. \ref{fig:2d_ring} for model Fractal2.}
    \label{fig:2d_fractal2}
\end{figure}

\begin{figure}
    \centering
    \includegraphics[width=1.0\linewidth]{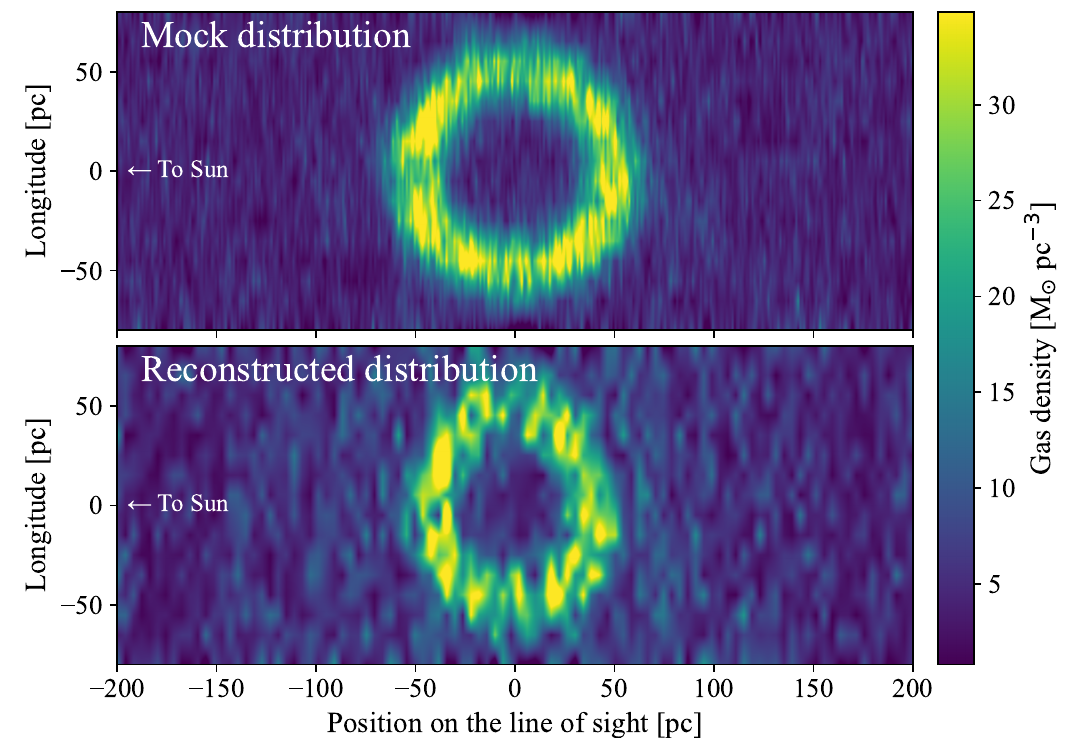}
    \caption{Same as Fig. \ref{fig:2d_ring} for model Fractal1.}
    \label{fig:2d_fractal1}
\end{figure}

Figs. \ref{fig:2d_fractal2} and \ref{fig:2d_fractal1} show the 2D maps. As for Fractal3 the shape and the intensity of the gas peaks are recovered.

\begin{figure}
    \centering
    \includegraphics[width=1.0\linewidth]{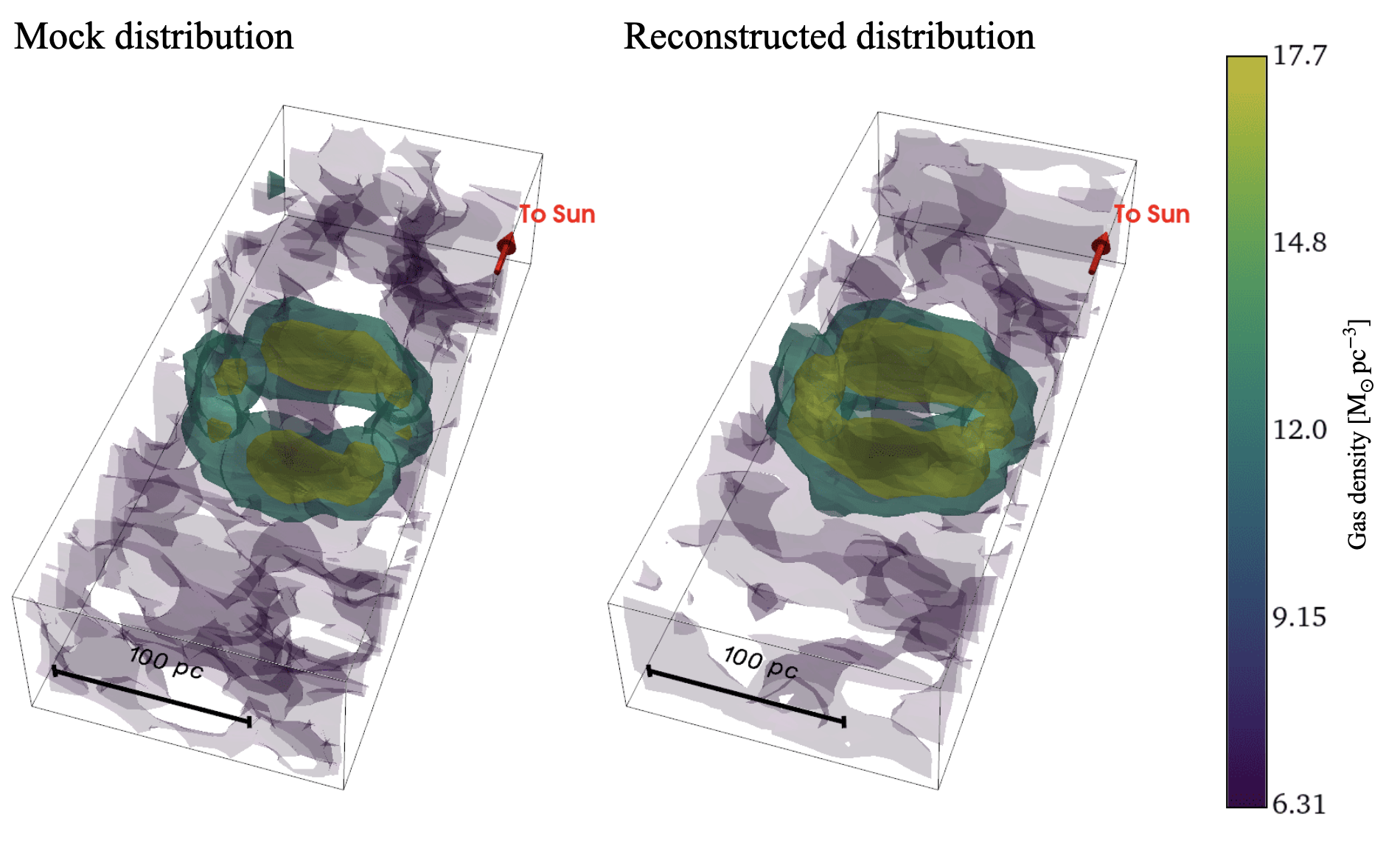}
    \caption{Same as Fig. \ref{fig:3d_ring} for model Fractal2.}
    \label{fig:3d_fractal2}
\end{figure}

\begin{figure}
    \centering
    \includegraphics[width=1.0\linewidth]{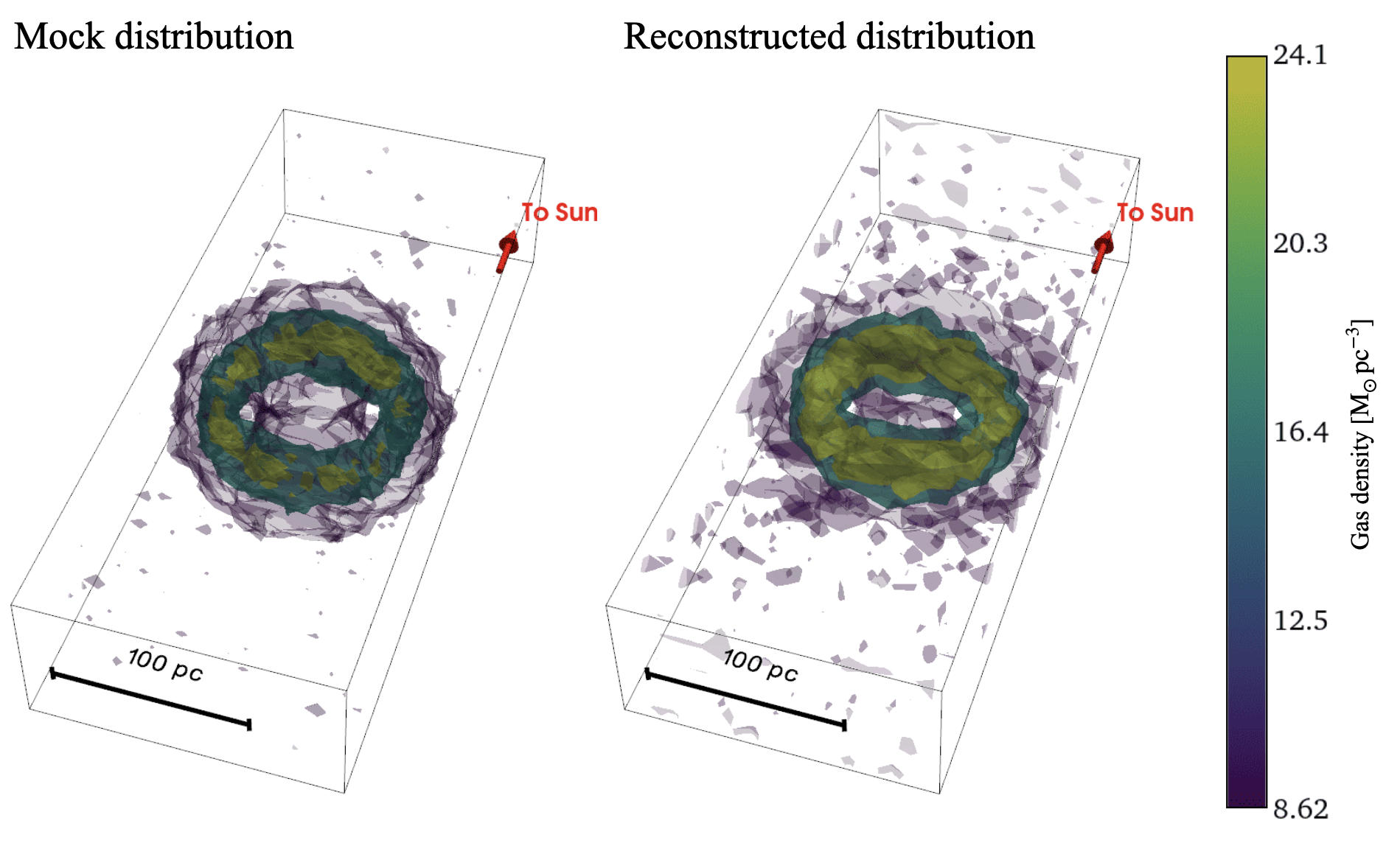}
    \caption{Same as Fig. \ref{fig:3d_ring} for model Fractal1.}
    \label{fig:3d_fractal1}
\end{figure}

Figs. \ref{fig:3d_fractal2} and \ref{fig:3d_fractal1} show the 3D maps. As for Fig.~\ref{fig:3d_fractal3} the 3D geometry is reconstructed correctly as well as the density of the different structures.

\end{appendix}
\end{document}